\documentclass{article}

\usepackage[text={173 mm,25cm},centering,letterpaper]{geometry}
\usepackage{amsmath, amsthm, amssymb}
\usepackage{graphicx}
\usepackage[margin = 10pt, font=small, labelfont=bf, labelsep = endash]{caption}
\usepackage[pdfpagemode=UseOutlines, hidelinks]{hyperref}
\usepackage{subcaption}
\usepackage[utf8]{inputenc}

\begin{document}

\title{Partial tipping in a spatially heterogeneous world}

\author{Robbin Bastiaansen\thanks{Institute for Marine and Atmospheric research Utrecht, Department of Physics, Utrecht University, Utrecht, The Netherlands (r.bastiaansen@uu.nl, h.a.dijkstra@uu.nl, a.s.vonderheydt@uu.nl)} , Henk A. Dijkstra\footnotemark[1] \thanks{Centre for Complex System Studies, Department of Physics, Utrecht University, Utrecht, The Netherlands} , Anna S. von der Heydt\footnotemark[1] \footnotemark[2] }

\maketitle

\begin{abstract}
    Many climate subsystems are thought to be susceptible to tipping -- and some might be close to a tipping point. The general belief and intuition, based on simple conceptual models of tipping elements, is that tipping leads to reorganization of the full (sub)system. Here, we explore tipping in conceptual, but spatially extended and spatially heterogenous models. These are extensions of conceptual models taken from all sorts of climate system components on multiple spatial scales. By analysis of the bifurcation structure of such systems, special stable equilibrium states are revealed: coexistence states with part of the spatial domain in one state, and part in another, with a spatial interface between these regions. These coexistence states critically depend on the size and the spatial heterogeneity of the (sub)system. In particular, in these systems a tipping point might lead to a partial tipping of the full (sub)system, in which only part of the spatial domain undergoes reorganization, limiting the impact of these events on the system's functioning.
\end{abstract}

\section{Introduction}

Many Earth system components and ecosystems have been shown to exhibit tipping~\cite{lenton2013environmental, scheffer2001catastrophic, holling1973resilience, drijfhout2015catalogue, lenton2008tipping}: when a tiny change in environmental conditions or parameters leads to a critical shift towards an alternative state that might have completely different functioning. For instance, the Amazonian rainforest that might disappear~\cite{hirota2011global, staver2011global}, desertification~\cite{rietkerk1997alternate, rietkerk1997site}, a restructuring of the Atlantic meridional overturning circulation~\cite{stocker1991rapid, lohmann2021risk}, collapses of ice sheets~\cite{garbe2020hysteresis, pattyn2020uncertain, huybrechts1999dynamic}, turbidity in shallow lakes~\cite{scheffer1993alternative}, amongst many others. Even on a planetary scale, tipping might have happened~\cite{lenton2013origin}, and is hypothesised to be possible in the (near) future~\cite{rockstrom2009safe, steffen2018trajectories, barnosky2012approaching}.

Typically, tipping is illustrated and explained using simple, conceptual low-dimensional models, that have two alternative states and that can tip between them as climatic conditions change~\cite{lenton2013environmental, holling1973resilience, may1977thresholds, scheffer2001catastrophic}. In more complex, more detailed high-dimensional models and in real-life data tipping is, however, often not as clear and pronounced~\cite{drijfhout2015catalogue, vannes2005, lenton2008tipping}. Tipping from one state to a completely differently structured state is hardly ever observed. Instead, partial restructurings occur more often. For instance, (large) parts of an ice sheet melt, instead of the whole sheet melting in one single tipping event~\cite{rosier2021tipping}.

This suggest that low- and high-dimensional models behave differently. It could be that high-dimensional models are tuned for stability too much, suppressing tipping behaviour~\cite{valdes2011built}. It could also be that the low-dimensional models are too restrictive in the number of physical processes, thereby exaggerating tipping behaviour~\cite{rietkerk2021evasion, bastiaansen2020effect}. At least, the most simple models really only allow for two alternative states and nothing more. Adding complexity to these leads to more response options for the system, which might lead to less severe tipping events. For instance, adding more boxes to a box model~\cite{gildor2001physical, alkhayuon2019basin}, or incorporating spatial effects~\cite{vannes2005, bastiaansen2020effect}.

\begin{figure}[p]
    \centering
        \begin{subfigure}[t]{0.45\textwidth}
            \centering
            \includegraphics[width=\textwidth]{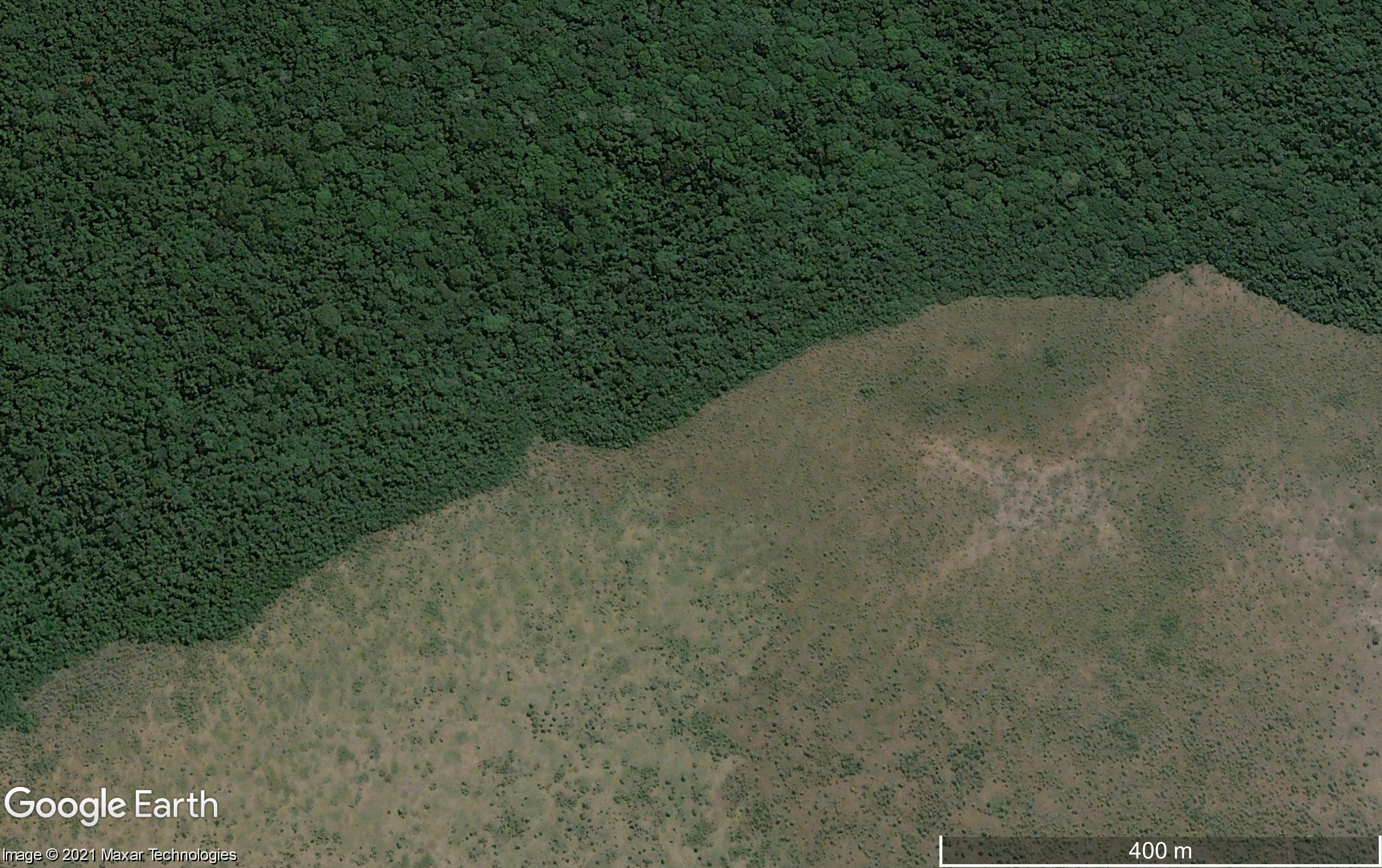}
            \caption{}
        \end{subfigure}
        \begin{subfigure}[t]{0.45\textwidth}
            \centering
            \includegraphics[width=\textwidth]{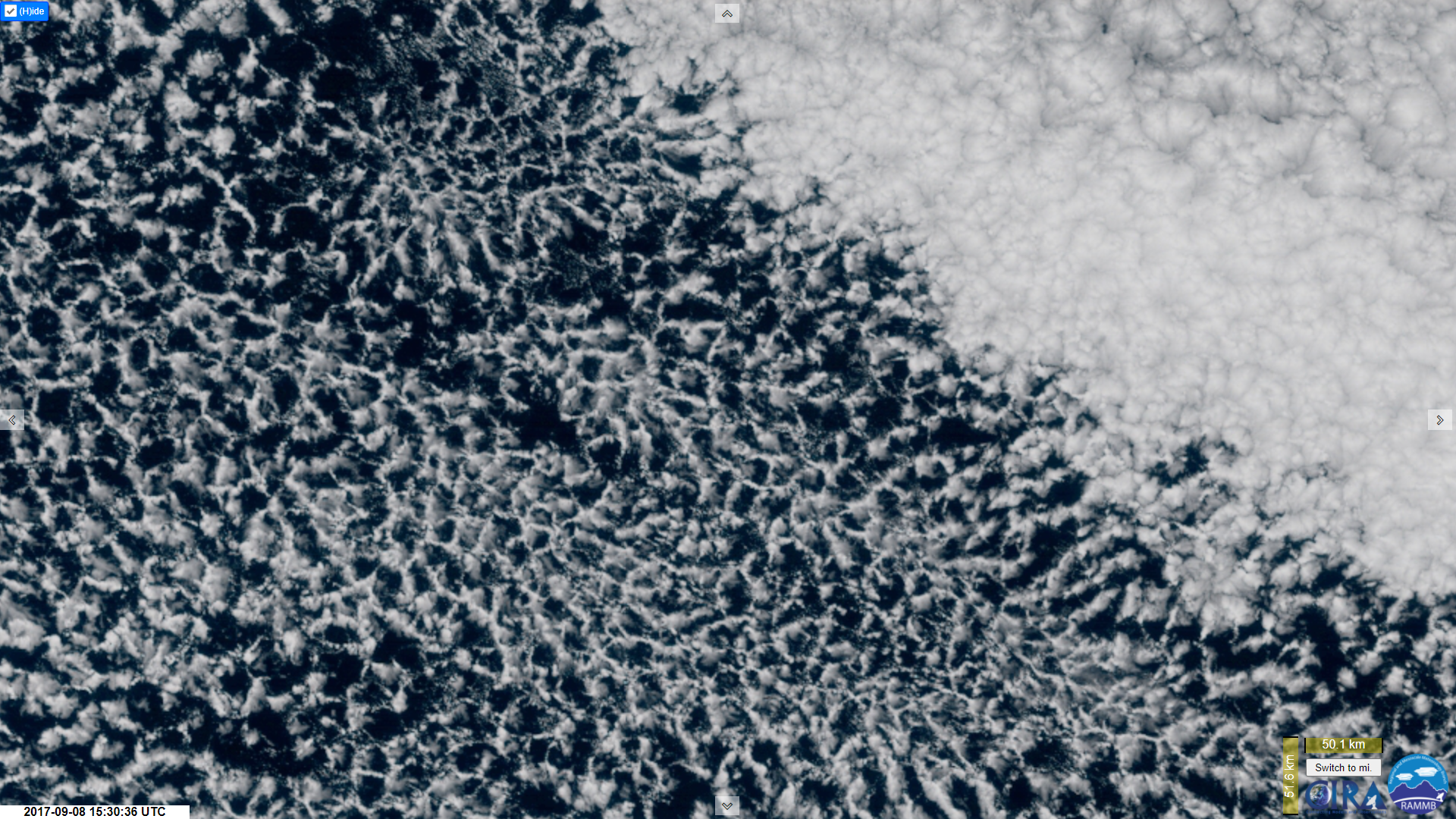}
            \caption{}
        \end{subfigure}
        \begin{subfigure}[t]{0.45\textwidth}
            \centering
            \includegraphics[width=\textwidth]{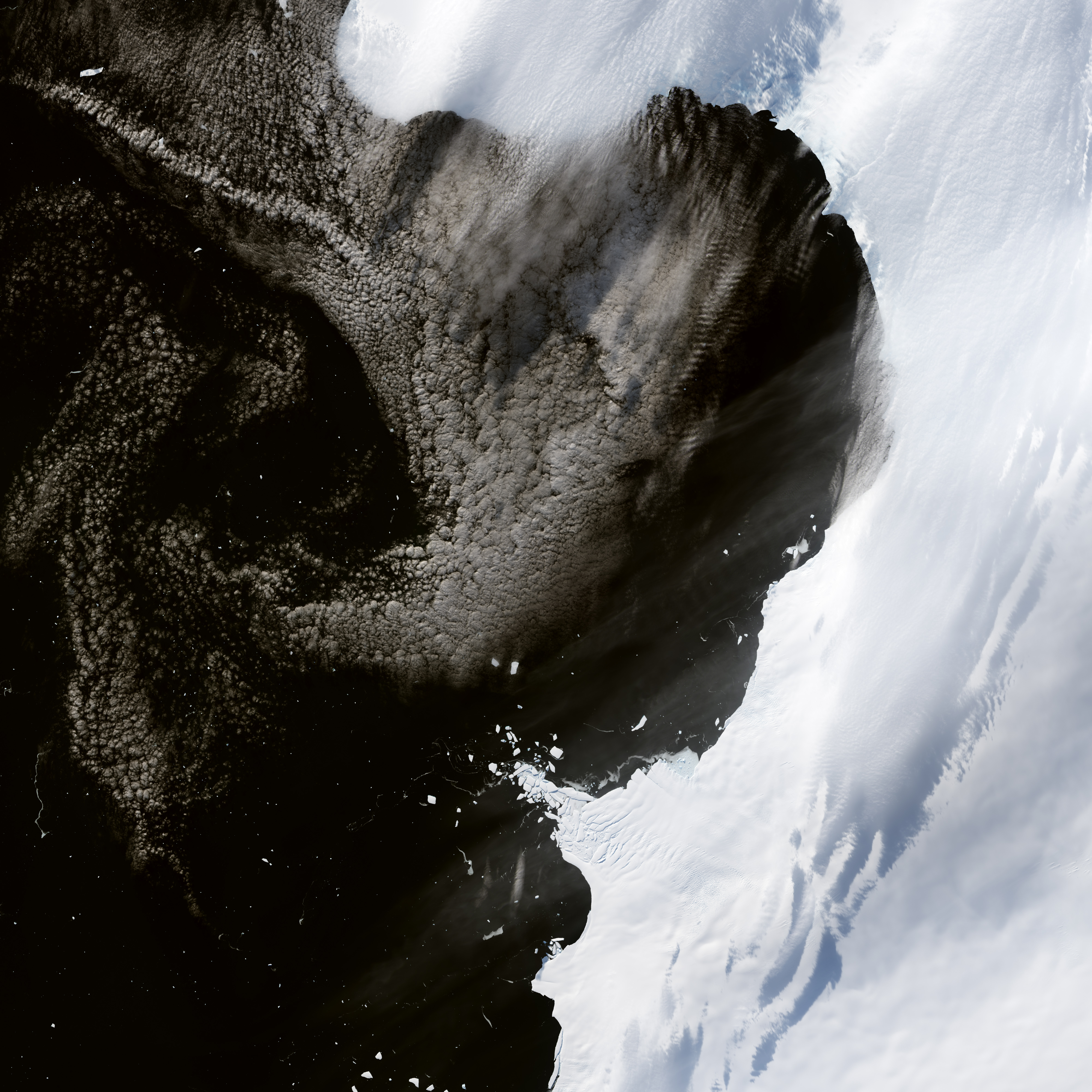}
            \caption{}
        \end{subfigure}
        \begin{subfigure}[t]{0.45\textwidth}
            \centering
            \includegraphics[width=\textwidth]{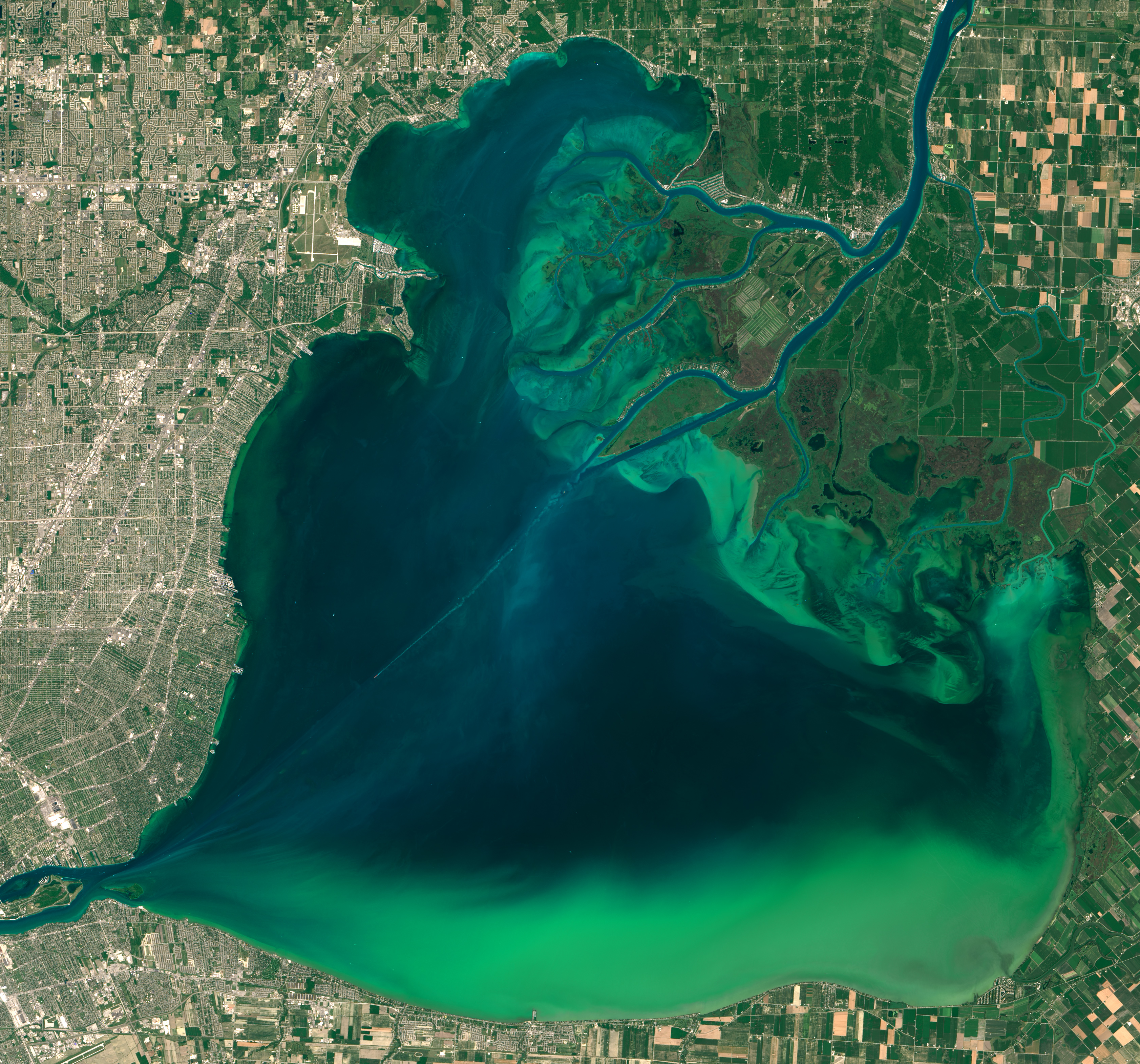}
            \caption{}
        \end{subfigure}
    \caption{Examples of visibly observable coexistence states and the spatial interfaces between the different states in real systems. (a) Spatial interface between tropical forest and savanna ecosystems (Google Earth image; Gabon, 1°19’47.15”S, 13°52’48.66'’E). (b) Spatial Interface between two types of stratocumulus clouds (RAMMB/CIRA SLIDER~\cite{Micke2018}; 9.45°S, 73.62°W, 2017-09-08 16:30:36 UTC). (c) Spatial interface between sea-ice and water in the Eltanin Bay in the Bellinghausen Sea (73°43'S, 83°49'W,  March 2, 2015. Image courtesy of NASA's Earth Observatory, NASA). (d) Algae bloom in part of Lake St. Clair on July 28, 2015 (Image courtesy of NASA's Earth Observatory, NASA).}
    \label{fig:real_examples}
\end{figure}

In this paper, we investigate the behaviour of conceptual models when spatial effects are incorporated: spatial transport and spatial heterogeneity. This setting has received only little attention in the literature~\cite{vannes2005, scheffer2012anticipating} and a thorough theoretical understanding of such systems is still lacking, despite their omnipresence~\cite{scheffer2003catastrophic}. In such models additional stable states called \emph{co-existence states} can emerge, in which part of the domain resides in one state and the rest in another state, with a spatial interface separating these regions~\cite{vannes2005, rietkerk2021evasion} -- see Figure~\ref{fig:real_examples} for real-life examples. Consequently, in these systems \emph{partial transitions} can occur in which only in part of the spatial domain the system changes state, providing a more subtle tipping pathway compared to the classic tipping example systems.

The rest of this paper is structured as follows. In section~\ref{sec:theory}, we explain theoretically the effect of spatial transport and spatial heterogeneity on tipping behaviour. We focus on the possibility of coexistence states and the different bifurcation diagrams these systems can have depending on the spatial heterogeneity. Then, in section~\ref{sec:examples}, we consider several examples of climate subsystems on different spatial scales, for which this theory might be relevant. Finally, we end with a discussion in section~\ref{sec:discussion}.

\section{Theory}\label{sec:theory}

We consider the evolution of a (single) state variable $y$. Depending on the particular system of interest, this could be for example temperature or vegetation coverage. Ignoring spatial effects for a moment, the local dynamics of $y$ are described by an ordinary differential equation, i.e.
\begin{equation}
    \frac{dy}{dt} = f(y;\mu) =: \frac{\partial V}{\partial y}(y;\mu),
    \label{eq:local_model}
\end{equation}
where $f$ describes the evolution over time, depending on the value of the state variable $y$ and the value of a (bifurcation) parameter $\mu$, and $V$ is the associated potential function (also referred to as the stability landscape)~\cite{lenton2013environmental, holling1973resilience, scheffer2001catastrophic}.

\begin{figure}[p]
    \centering
        \begin{subfigure}[t]{0.32\textwidth}
            \centering
            \includegraphics[width=\textwidth]{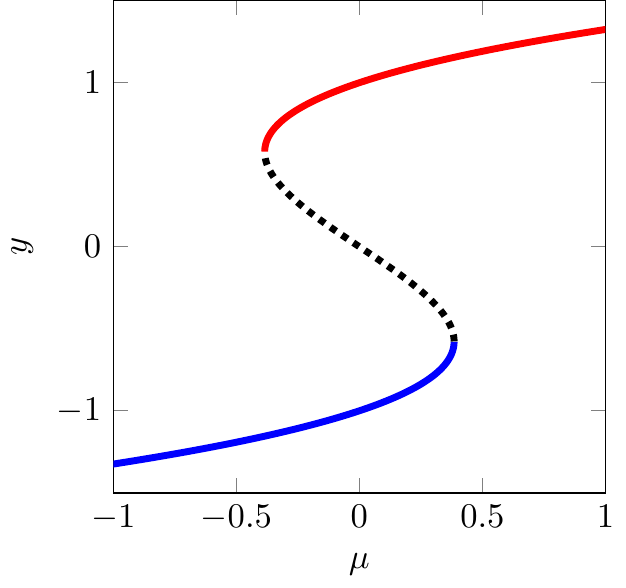}
            \caption{}
            \label{fig:bifurcation_diagram_0D}
        \end{subfigure}
        \begin{subfigure}[t]{0.32\textwidth}
            \centering
            \includegraphics[width=\textwidth]{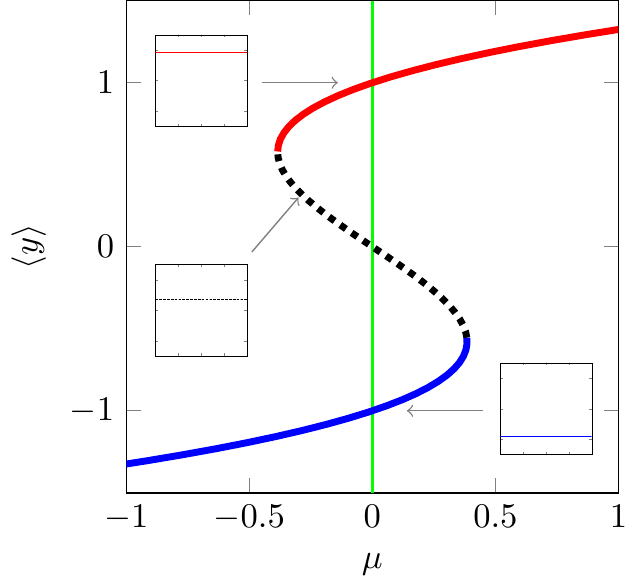}
            \caption{}
            \label{fig:bifurcation_diagram_1D_diffusion}
        \end{subfigure}
        \begin{subfigure}[t]{0.32\textwidth}
            \centering
            \includegraphics[width=\textwidth]{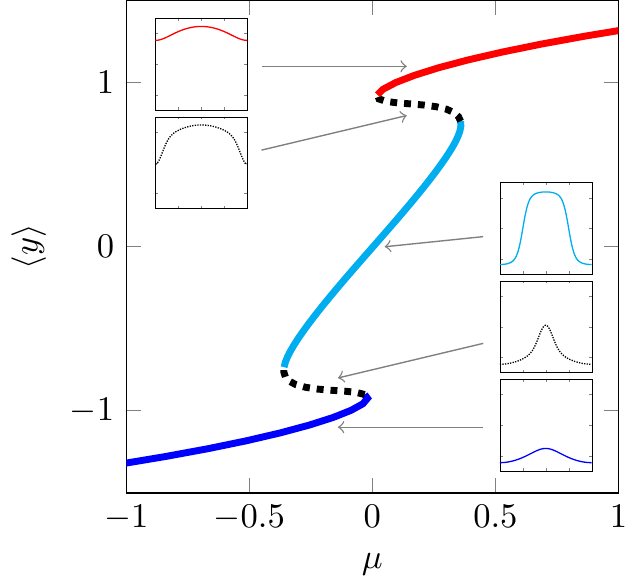}
            \caption{}
            \label{fig:bifurcation_diagram_1D}
        \end{subfigure}
    \caption{Bifurcation diagram for (a) a non-spatial model, (b) a spatially homogeneous model and (c) a spatially heterogeneous model. Solid lines denote stable equilibria and dotted lines unstable ones. The vertical axes in (b) and (c) indicate the spatial average of the state variable. Insets in (b) and (c) show the spatial structure of equilibria for the different branches. The green line in (b) indicates the Maxwell point of the bifurcation parameter. The cyan branch in (c) consists of stable coexistence states. For (c), the model used is $\frac{\partial y}{\partial t} = \frac{1}{100} \frac{\partial^2 y}{\partial x^2} + y(1-y^2) + \frac{1}{2}\cos(x\pi)$ on the domain $[-1,1]$ with Neumann boundary conditions; for (b) the spatially heterogeneous terms are absent and for (a) also the diffusion term is absent. }
    \label{fig:bifurcation_diagrams}
\end{figure}

The model (\ref{eq:local_model}) possesses tipping behaviour when there is a parameter region in which there is bistability of two different states -- say, a state $A$ and state $B$ -- with tipping points (saddle-node bifurcations) to the other state at both ends of this region. This leads to the prototypical `S'-curve bifurcation diagram as shown in Figure~\ref{fig:bifurcation_diagram_0D}. An explicit example of such system is the model~(\ref{eq:local_model}) with
\begin{equation}
    V(y;\mu) = \ \frac{y^2}{2} - \frac{y^4}{4} + \mu y.
\end{equation}
In the rest of this section, we detail how the addition of spatial effects to the model~(\ref{eq:local_model}) changes the bifurcation diagram.

\subsection{Spatial Transport}
We now consider the evolution of the state variable $y$ on a spatial domain with coordinate $x$. In addition to the local dynamics, spatial transport between locations now also plays a role. The most simple way to implement spatial transport is to add (linear) diffusion to the model~(\ref{eq:local_model}). Thus, a new spatial model describing the evolution of $y$ over time and space is given by the partial differential equation
\begin{equation}
    \frac{\partial y}{\partial t} = D \Delta y + f(y;\mu) = D \Delta y + \frac{\partial V}{\partial y}(y;\mu),
    \label{eq:spatial_model_diffusion}
\end{equation}
where $D$ is the diffusion coefficient and $\Delta$ the spatial Laplacian in $x$.

Because this type of model can be used to model phase segregation dynamics \cite{allen1972ground, bray2002theory}, its dynamics are well understood and the literature is quite extensive, including rigorous mathematical proofs \cite{carr1989metastable, sandstede2002stability}. First, spatially uniform states exist in which the whole domain is in the same state (either state $A$ or $B$). These are the natural spatially extended versions of the states in the local model~(\ref{eq:local_model}). However, next to these, there are also coexistence states, in which parts of space reside in state $A$ and the rest in state $B$, with a spatial interface (or `front' in mathematical jargon) connecting these regions. These interfaces can be spatially localized, or extend over a large space depending on the strength of the spatial transport.

The dynamics of these coexistence states is intricate~\cite{carr1989metastable, pismen2006patterns}. In short: the interfaces typically migrate slowly over space depending on the distance to other interfaces and the domain boundary, but there are sometimes fast coalescence events when interfaces meet and separate regions merge. On the long term, a maximum of one interface can persist in the domain (the rest is slow transient behaviour). Such single interface moves with a speed that is proportional to the difference in potential between states $A$ and $B$, i.e. $V(A;\mu) - V(B;\mu)$. So, only in the degenerate case in which the potentials are equal, this leads to an additional equilibrium state of~(\ref{eq:spatial_model_diffusion}). The specific parameter value for which the potentials are equal is called the \emph{Maxwell point} of the system. This means that the bifurcation diagram does not change much compared to the non-spatial system; see Figure~\ref{fig:bifurcation_diagram_1D_diffusion}. However, the transient dynamics occurring after a tipping point has been crossed can be different in such systems: this can now occur via an invasion front which replaces one state with the other, which might lead to a slow tipping of the full system~\cite{zelnik2018regime, bel2012gradual}.

\subsection{Spatial Heterogeneity}

The previous model~(\ref{eq:spatial_model_diffusion}) assumes that the local dynamics are the same throughout the whole domain, i.e. the model is spatially homogeneous. Reality is, however, spatially heterogeneous: local dynamics differ from location to location, for example due to topographical features or human activity. Hence, it is more realistic to consider a spatially extended model in which the local dynamics explicitly depends on the location $x$:
\begin{equation}
    \frac{\partial y}{\partial t} = D \Delta y + f(y, x; \mu) = D \Delta y + \frac{\partial V}{\partial y}(y, x; \mu).
    \label{eq:spatial_model}
\end{equation}
We will assume that the spatial variation is not too much locally, i.e. $\frac{\partial f}{\partial x}(y, x; \mu)$ does not get very large compared to diffusion strength (as this would, potentially, give rise to different dynamics than described in this paper).

The presence of spatial heterogeneity in~(\ref{eq:spatial_model}) does change the dynamics of the interfaces. Their movement now depends on the \emph{local} difference between the potentials of state $A$ and $B$. Because of this, the Maxwell point is different for different locations in space, making it no longer a degenerate case. Furthermore, coexistence states with multiple interfaces can now be equilibrium states of the model system.

\begin{figure}[p]
    \centering
     \begin{subfigure}[t]{0.32\textwidth}
            \centering
            \includegraphics[width=\textwidth]{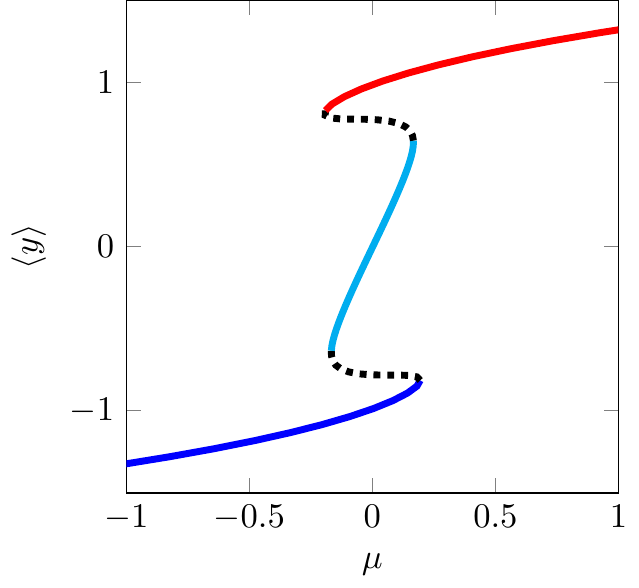}
            \caption{}
        \end{subfigure}
     \begin{subfigure}[t]{0.32\textwidth}
            \centering
            \includegraphics[width=\textwidth]{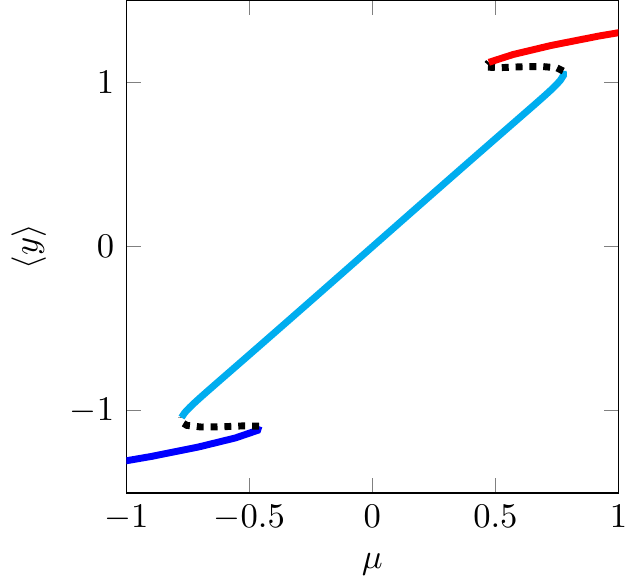}
            \caption{}
        \end{subfigure}
     \begin{subfigure}[t]{0.32\textwidth}
            \centering
            \includegraphics[width=\textwidth]{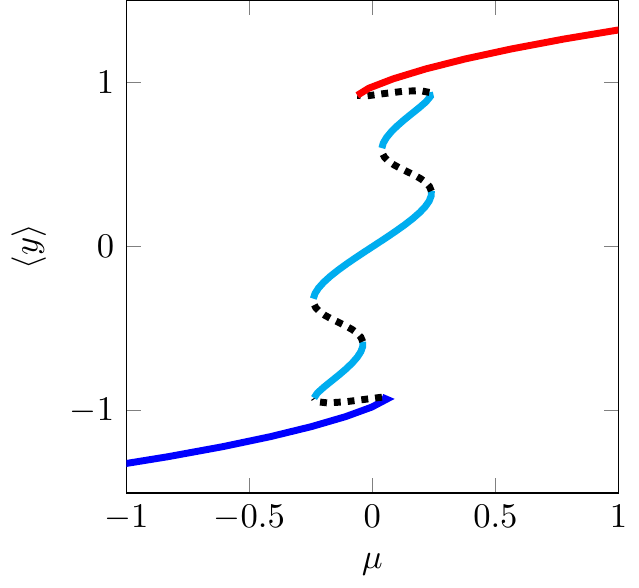}
            \caption{}
        \end{subfigure}
        
     \begin{subfigure}[t]{0.32\textwidth}
            \centering
            \includegraphics[width=\textwidth]{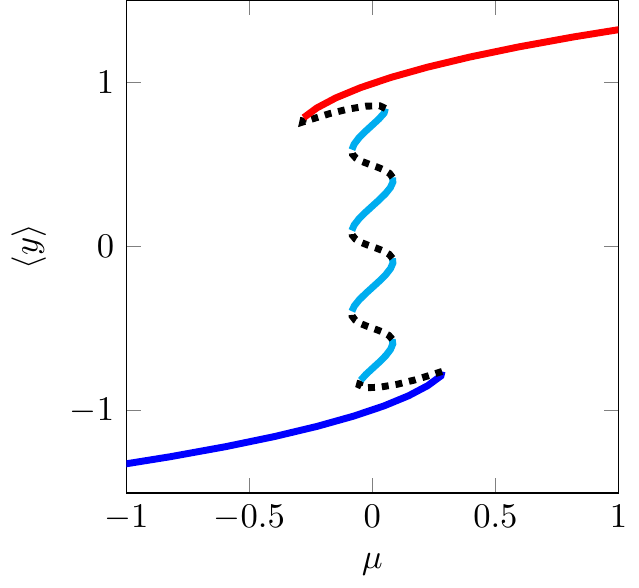}
            \caption{}
        \end{subfigure}
     \begin{subfigure}[t]{0.32\textwidth}
            \centering
            \includegraphics[width=\textwidth]{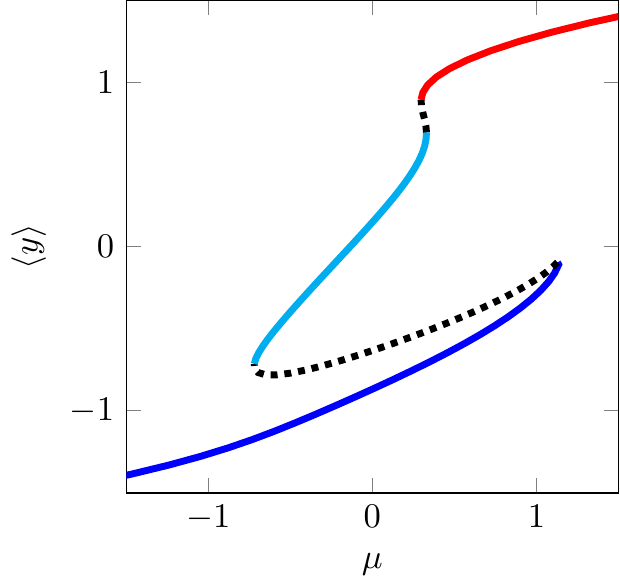}
            \caption{}
        \end{subfigure}
     \begin{subfigure}[t]{0.32\textwidth}
            \centering
            \includegraphics[width=\textwidth]{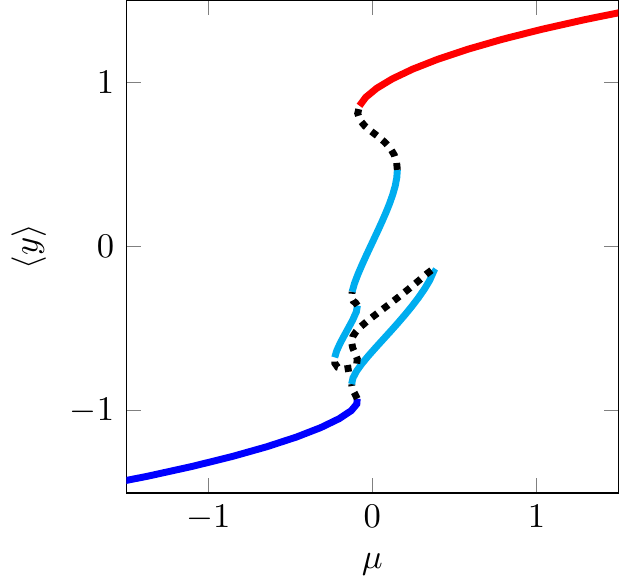}
            \caption{}
        \end{subfigure}

    \caption{Showcase of some variations of the bifurcation diagram for spatially heterogeneous models with different spatial heterogeneities. The model used is $\frac{\partial y}{\partial t} = \frac{1}{100} \frac{\partial^2 y}{\partial x^2} + y (1 - y^2) + g(x,y)$ on the domain $[-1,1]$ with Neumann boundary conditions. In (a) $g(x,y) = \frac{1}{4} \cos(\pi x)$, (b) $g(x,y) = x$, (c) $g(x,y) = \frac{1}{2}\cos(\pi x)\sin(\frac{3}{2}\pi x)$, (d) $g(x,y) = \cos(\pi x)\cos(2\pi x)\sin(\pi x)$, (e) $g(x,y) = 2 \cos(\pi x) y + \cos(2 \pi x)$ and (f) $g(x,y) = \frac{1}{2}\cos(\pi x)y + \frac{1}{2}\cos(2\pi x)$. Bifurcation diagrams might not show all possible branches, because isolated branches might exist.}
    \label{fig:bifurcation_diagram_variations}
\end{figure}

A typical bifurcation diagram is given in Figure~\ref{fig:bifurcation_diagram_1D}, which is different compared to the classic one (Figure~\ref{fig:bifurcation_diagram_0D}) because of the additional branch of stable coexistence states. In this case, the crossing of a tipping point does not necessarily lead to a tipping of the complete system, but can lead to a \emph{partial tipping} of the system in which only part of the spatial domain undergoes a transition to the alternative state. The precise structure of the bifurcation diagram depends on the specific heterogenity of the system. A few possible variations are shown in Figure~\ref{fig:bifurcation_diagram_variations}, illustrating the potential complexity of the bifurcation diagram for spatially heterogeneous systems.

A concise description of the mathematical theory and analysis of coexistence states can be found in~\ref{sec:app_math_coexistence}.

\begin{figure}
    \centering
    \includegraphics{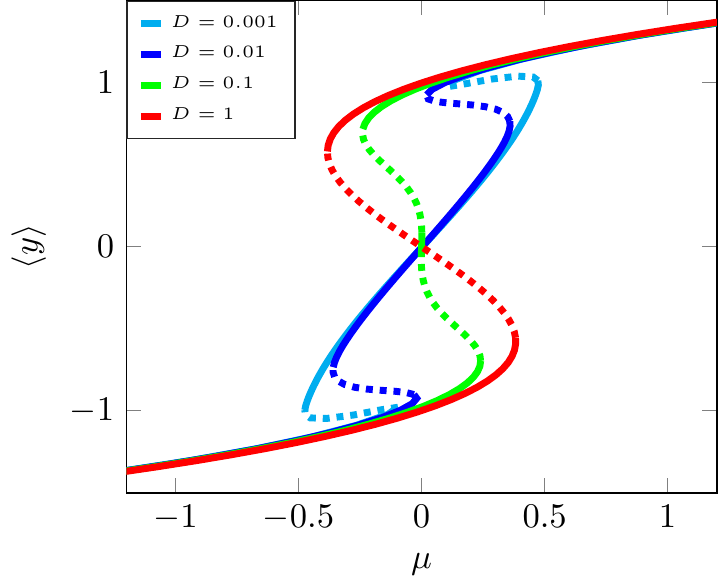}
    \caption{Bifurcation diagrams for different diffusion strengths, but the same spatial heterogeneity and domain size, demonstrating the change of structure of the diagram as the domain becomes too small to fit in the spatial interfaces (as more space is required when $D$ is larger). The model used is $\frac{\partial y}{\partial t} = D \frac{\partial^2 y}{\partial x^2} + y (1 - y^2) + \mu + \frac{1}{2} \cos(\pi x)$ on $[-1,1]$ with no-flux boundaries.}
    \label{fig:bifurcation_diagram_lengths}
\end{figure}

The above described spatial interfaces between regions in different states can only emerge and persist if there is enough space available for them in the system. Hence, if the spatial domain is too confined, the above described coexistence states do not fit and the branch of coexistence states is absent in the bifurcation diagram (see Figure~\ref{fig:bifurcation_diagram_lengths}). The precise minimum required length varies per model and depends on the typical length scale associated with the spatial transport. For instance, if the diffusion coefficient $D$ in~(\ref{eq:spatial_model}) is increased, a larger domain is needed to facilitate the coexistence states. See also \ref{sec:small_domain_limit} for a mathematical treatment of the small domain limit.

The form of equation~(\ref{eq:spatial_model}) is chosen for simplicity of presentation. Coexistence states can also occur in other models that for example have a diffusion coefficient that depends on space or have a different spatial transport mechanism altogether. Similar behaviour as described above can at least be expected from other spatially heterogeneous Lagrangian or gradient systems.

\section{Earth system and ecosystem example conceptual models}\label{sec:examples}

The above described theory is quite generic, and there are only few assumptions. Hence, if a simple conceptual model only considers local dynamics and captures tipping behaviour, a spatially extended and heterogeneous model will behave as described above. In particular, such systems might have additional coexistence states, and partial tipping events in which only part of the domain tips. Many systems, and many previously proposed conceptual models, fit into this description. To illustrate this, next we give some examples of systems for which this theory might be relevant. Hereby we cover systems of many different spatial scales, ranging from global to more regional systems.

\begin{figure}[p]
    \centering

    \begin{subfigure}[t]{0.32\textwidth}
            \centering
            \includegraphics[width=\textwidth]{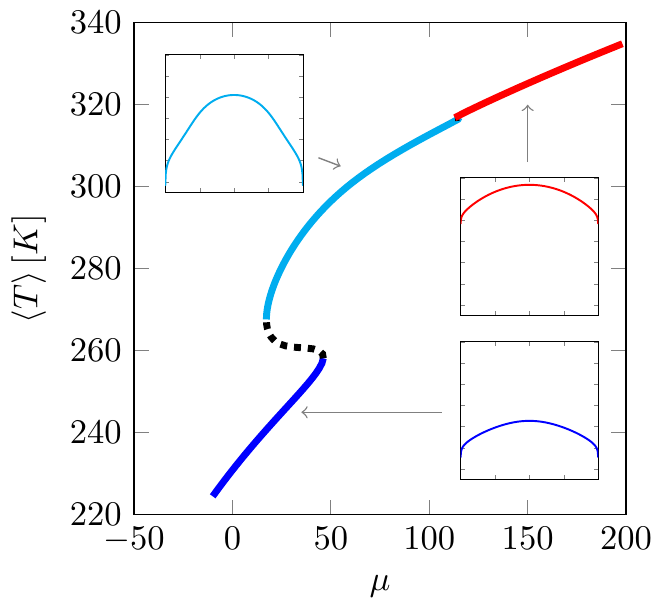}
            \caption{1D GEBM}
            \label{fig:BD-GEBM}
    \end{subfigure}
    \begin{subfigure}[t]{0.32\textwidth}
            \centering
            \includegraphics[width=\textwidth]{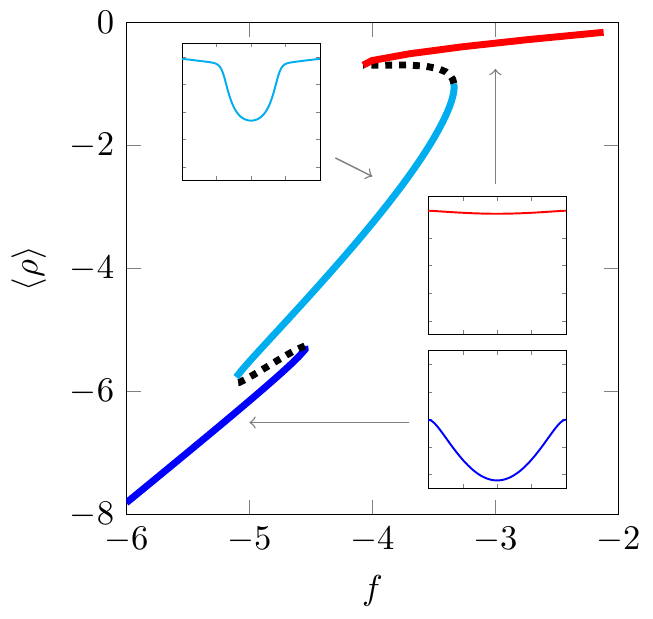}
            \caption{Ocean convection model}
            \label{fig:BD-ocean}
    \end{subfigure}
    \begin{subfigure}[t]{0.32\textwidth}
            \centering
            \includegraphics[width=\textwidth]{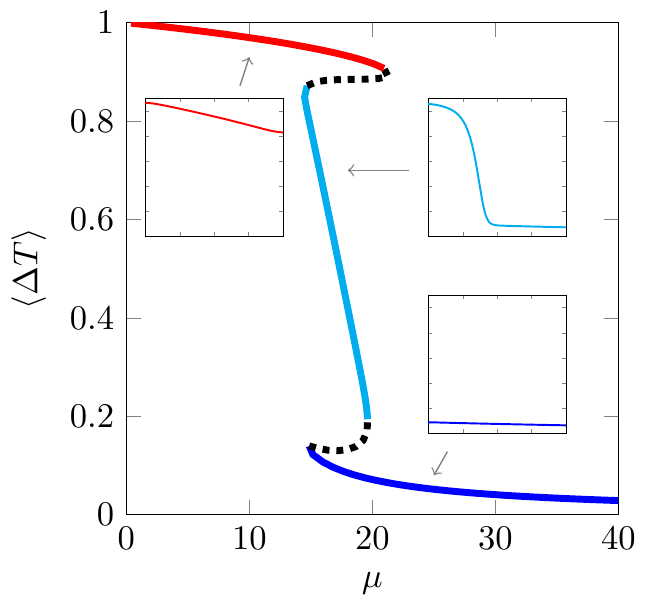}
            \caption{ABL}
            \label{fig:BD-ABL}
    \end{subfigure}

    \begin{subfigure}[t]{0.32\textwidth}
            \centering
            \includegraphics[width=\textwidth]{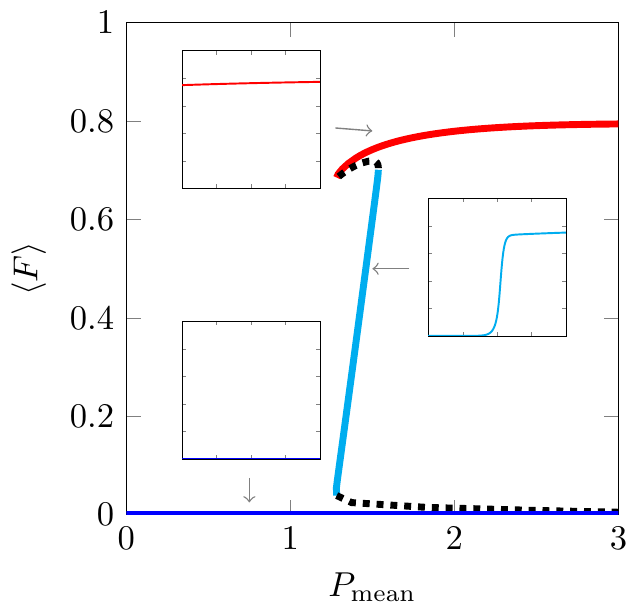}
            \caption{Tropical forest model}
            \label{fig:BD-tropicalforest}
    \end{subfigure}
    \begin{subfigure}[t]{0.32\textwidth}
            \centering
            \includegraphics[width=\textwidth]{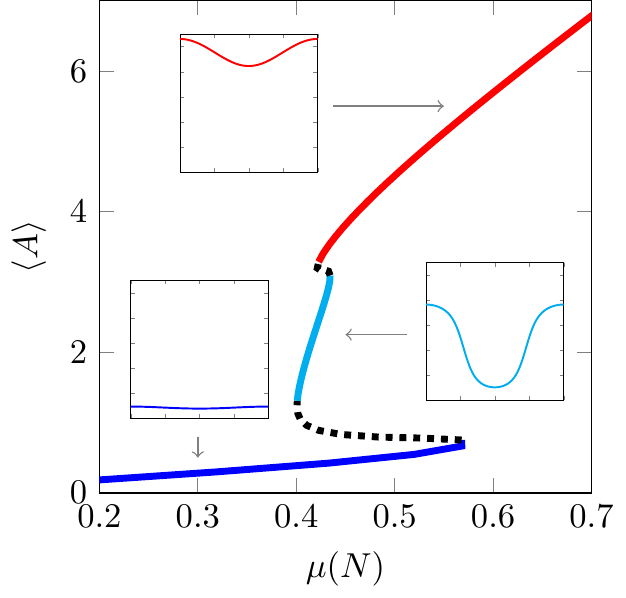}
            \caption{Shallow Lakes}
            \label{fig:BD-lakes}
    \end{subfigure}

    \caption{Bifurcation diagrams for the example systems in section~\ref{sec:examples}. In all figures, the red and blue branches denote equilibria in which the whole system is in the same state. The cyan branch denotes the stable coexistence states, in which part of the spatial domain is one state and the rest in the other. The insets show the spatial structure of solutions along the stable branches. (a) Global Energy Balance Model (GEBM). (b) Ocean convection model. (c) Atmospheric boundary layer model. (d) Tropical forest model. (e) Model for turbidity in shallow lakes. }
    \label{fig:BD_examples}
\end{figure}

\subsection{Global Energy Balance Model}
On a planetary scale, an example of coexistence states can be found in global energy balance models. These describe the evolution of Earth's temperature by considering the energy flux at the top of the atmosphere. These type of models were first introduced by Budyko and Sellers~\cite{budyko1969effect, sellers1969global}, and have since served often as conceptual models to illustrate planetary climate shifts~\cite{ikeda1999study, north1981energy, ghil1976climate, bodai2015global, thorndike2012multiple, north1984small, north1975analytical, drazin1977branching, caldeira1992susceptibility, hoffman2002snowball, ashwin2020extreme}.

Here, we consider the following simple spatially one-dimensional variant:
\begin{equation}
    C_T \frac{\partial T}{\partial t} = Q_0(x) \left[1-\alpha(T)\right] - \varepsilon \sigma_0 T^4 + \mu + D \frac{\partial}{\partial x}\left[ \left(1-x^2\right)\frac{\partial T}{\partial x} \right].
    \label{eq:GEBM}
\end{equation}
This models the longitudinally averaged temperature $T$ (as function of $x = \sin(\theta)$ where $\theta$ is the latitude) by looking at the absorbed incoming solar radiation ($Q_0(x)\left[1-\alpha(T)\right]$), the outgoing Planck radiation ($-\varepsilon \sigma_0 T^4$), the meridional heat transport ($+ D \frac{\partial}{\partial x}\left[ \left(1-x^2\right)\frac{\partial T}{\partial x} \right]$), and the effect of atmospheric CO$_2$ on the energy budget ($+\mu$). More details, including parameter values, can be found in~\ref{sec:app_GEBM}.

By incorporating the ice-albedo feedback, and hence the temperature-dependency of albedo (ice, with a high albedo, can only exist if temperatures are low enough), it has been shown that the model can reside in an ice-free state or a fully ice-covered ``Snowball Earth'' state. When spatial effects are not modelled, these are the only states of the model. However, once spatial effects are incorporated, coexistence states are also found with ice in only part of the earth (i.e. the poles) and no ice elsewhere~\cite{ikeda1999study, north1981energy, ghil1976climate, bodai2015global, thorndike2012multiple, north1984small, north1975analytical, drazin1977branching, caldeira1992susceptibility, hoffman2002snowball}. Specifically, for such models, the meridional heat transport and the latitudinal dependency of incoming solar radiation can lead to coexistence states on a planetary scale. An example bifurcation diagram is given in Figure~\ref{fig:BD-GEBM} which shows the global mean temperature as function of the parameter $\mu$.

\subsection{Ocean Circulation}\label{sec:example_oceans}

In oceans, convection occurs at places where the ocean mixed-layer exchanges water with the deep ocean. This ocean convection plays an important role in the global ocean circulation. Using simple conceptual models, the possibility of bistability between a convective and a non-convective state has been found for a range of parameters~\cite{welander1982simple, lenderink1994variability, den2011spurious, vellinga1998multiple}. These results have been obtained in a box model, where temperature $T$ and salinity $S$ of the mixed-layer are dynamic variables, which change through exchange with (static) atmosphere and deep ocean boxes. Here, we extend the model by incorporation of one spatial dimension (representing the location in an ocean basin) via the addition of diffusion:
\begin{eqnarray}
    \frac{\partial T}{\partial t} & = D \frac{\partial^2 T}{\partial x^2} + k_T \left( T_A(x) - T \right) - \kappa(\Delta \rho) \left( T - T_0(x) \right); \label{eq:ocean_model_eq1}\\
        \frac{\partial S}{\partial t} & = D \frac{\partial^2 S}{\partial x^2} + k_S \left( S_A(x) - S \right) - \kappa(\Delta \rho) \left( S - S_0(x) \right), \label{eq:ocean_model_eq2}
\end{eqnarray}
where $\Delta \rho := \rho - \rho_0$ is the density difference between the mixed-layer and the deep ocean. Further, $T_A(x)$ and $S_A(x)$, respectively $T_0(x)$ and $S_0(x)$, are the space-dependent (reference) temperature and salinity of the atmosphere and deep ocean box, respectively. $k_T$ and $k_S$ give the rate of exchange with the atmosphere box and $\kappa(\Delta \rho)$ the rate of exchange with the deep ocean box. The latter is a function of $\Delta \rho$, which is a non-negative increasing function~\cite{welander1982simple}. That is, convection happens when the exchange rate $\kappa(\Delta \rho)$ is large, which typically occurs if the mixed-layer is heavy enough compared to the deep ocean.

To illustrate coexistence states with convection only at certain regions, we apply a series of further simplification on the equations (details in \ref{sec:app_ocean}). This leads to the following (scaled) equation that describes the evolution of the local density difference $\Delta \rho(x,t)$ between the mixed-layer and the deep ocean:
\begin{equation}
    \frac{\partial \Delta \rho}{\partial t} = D \frac{\partial^2 \Delta \rho}{\partial x^2} + k_T \left( \Delta \rho_A(x) - \Delta \rho \right) - \kappa(\Delta \rho) \Delta \rho + D \frac{\partial^2 \rho_0(x)}{\partial x^2}.
\end{equation}
Depending on the functional form of the exchange rate $\kappa(\Delta \rho)$ between the mixed-layer and the deep ocean, certain parameter ranges can allow for bistability and coexistence states between convection and no convection. Not striving for the most realistic description, we take a continuous approximation of a step function similar to~\cite{welander1982simple}. Specifically, we take $\kappa(\Delta \rho) = \frac{\bar{\kappa}}{2} \left[ 1 + \tanh\left(\Delta \rho-\Delta \rho_\mathrm{ref}\right)\right]$ which goes from $0$ to $\bar{\kappa}$ and starts to increase around $\Delta \rho_\mathrm{ref}$. 

To illustrate this model, we take a spatially heterogeneous atmospheric reference density $\Delta \rho_A(x) = 2 + f \left(1 + \cos\left[\frac{\pi x}{2}\right]\right)$, where $f$ acts as a bifurcation parameter. In ocean systems, such kind of spatial heterogeneity could for instance be caused by local differences in freshwater fluxes. In figure~\ref{fig:BD-ocean}, an example bifurcation diagram for this model is given, showing the possibility of coexistence states in large enough ocean basins.

\subsection{Atmospheric Circulation}

In the atmospheric circulation, there also seems to be the possibility of coexistence states. As a simple example, we look at the dynamics of the atmospheric boundary layer, in which heat exchange with the surface takes place. This boundary layer can be in a fully turbulent state or a quiescent, quasi-laminar state depending on the temperature difference between the boundary layer and the soil. In cold regions, or during cold nights, the layer is stable stratified, but at higher temperatures surface warming leads to a convective boundary layer.

Bistability and tipping between these two states has been explained using simple energy balance models~\cite{kaiser2020detecting, vandewiel2017regime}. These describe the near-surface inversion strength $\Delta T$, defined roughly as the boundary layer's temperature minus the soil temperature, which evolves according to the net imbalance of the energy fluxes in the layer~\cite{vandewiel2017regime}. Here, we have extended such model by adding spatial heat re-distribution as a diffusive process. The (scaled) model is given by
\begin{equation}\label{eq:atmosphere_model}
    \frac{\partial \Delta T}{\partial t} = D \frac{\partial^2 \Delta T }{\partial x^2} + Q(x) - \lambda \Delta T - C(x) \Delta T  e^{-2 \alpha \Delta T }.
\end{equation}
Here, $Q(x) - \lambda \Delta T$ is the linearized net long-wave radiation, and $-C(x)\Delta T e^{-2 \alpha \left(\Delta T \right)}$ is the turbulent sensible heat flux. The location $x$ can be interpreted as a local coordinate of a larger region on Earth.

To demonstrate the effect of spatial heterogeneity in this model, we take $
\lambda = 1$, $\alpha = 3 $, $Q(x) = 1 - 0.1 x$ and $C(x) = 0.1 x + \mu$, modelling for example spatial differences in soil temperature or wind speeds (the latter influences the turbulent sensible heat flux); $\mu$ acts here as a bifurcation parameter, indicating changes in wind speed. For these choices, an example bifurcation diagram is given in Figure~\ref{fig:BD-ABL}, which indicates the possibility of coexistence states with regional turbulence.

\subsection{Ecosystems}

In many ecological systems, bistability and transitions between alternative states have been found. For instance, in drylands, bistability between vegetation and bare soil occurs~\cite{rietkerk2004self}. On larger scales, the same environmental conditions can even support multiple ecosystem types, such as grasslands, savannas and tropical forests in the Tropics~\cite{staver2011global, aleman2020floristic}. In all of these systems, spatial effects are present: species disperse over space (e.g. via seed dispersal) and environmental conditions are spatially heterogeneous (e.g. a rainfall gradient). Hence, they can exhibit coexistence states in which different ecosystem types are separated by a local interface, and can show partial tipping.

Especially in models of drylands, many patterned states have been analyzed, including coexistence states~\cite{bel2012gradual, bastiaansen2019stable}, although spatial heterogeneities are not often taken into account, with few exceptions~\cite{bastiaansen2020pulse, gandhi2018topographic}. However, the models considered are typically more complicated than~(\ref{eq:spatial_model}) and more advanced mathematical techniques are used to analyze them that are beyond the scope of this article. Hence, we illustrate the coexistence patterns in ecosystems using a different example.

Recently, the bistability between tropical forests and grasslands in a heterogeneous environment has been studied~\cite{wuyts2019tropical}. In a spatially extended (rescaled) version of the model in~\cite{staver2012integrating}, the local fraction of fire-prone forest trees at is given by $F$ and its evolution is modelled as
\begin{equation}
    \frac{\partial F}{\partial t} = D \frac{\partial^2 F}{\partial x^2} + r(P) (1 - F) F - m(P) F - F f(F;P),
    \label{eq:tropical_forest}
\end{equation}
with the coordinate $x$ representing e.g. a cross-section through a region. Here, $r(P)$ is the (logistic) growth rate depending on local precipitation $P(x)$. $m(P)$ is the precipitation-dependent natural mortality rate. $f(F;P)$ is the mortality due to forest fires, which depends both on the local precipitation $P(x)$ and the local tree coverage $F$ (dry trees burn easier and fire spreads easier when there are less trees and more herbaceous vegetation). Finally, seed dispersal is modelled as diffusion. Details of the functional forms and parameter values are given in \ref{sec:app_tropicalforest}.

The spatially heterogeneous term in this example is the local precipitation $P(x)$. We have modelled this as a linear precipitation gradient, and model climate change via the mean precipitation parameter $P_\mathrm{mean}$. An example bifurcation diagram is given in Figure~\ref{fig:BD-tropicalforest}. Here, the red branch has solutions with non-zero forest tree fraction $F$ everywhere, the blue branch corresponds to a no-forest state and the cyan branch to coexistence states with forest trees in part of the domain. In this diagram, the blue branch is attracting for all parameter values, and does not connect to the other branches via a saddle-node bifurcation (the lower unstable branch only connects to the blue branch in the limit $P_\mathrm{mean} \rightarrow \infty$).

\subsection{Shallow lakes}

On even smaller spatial scales, coexistence states could also play a role. As an example, we look at the turbidity of shallow lakes~\cite{scheffer1993alternative}, which has been used as one of the prime examples of critical shifts in ecological systems. In these lakes, turbidity is closely related to the algae concentration $A$, which can change depending on the amount of nutrients in the lake. If there is a lot of nutrients, algae concentration is high and the lake is turbid. For lower nutrient concentrations, there are few algae and the lake is clear. Transitions between these states can take place if the nutrient concentration changes.

In~\cite{scheffer1993alternative}, a simple non-spatial model is given for the evolution of algae $A$, depending on the nutrient concentration. Here, we have added spatial transport as a diffusive process to obtain the following (rescaled) spatially explicit model that models the local algae fraction over e.g. a cross-section of a shallow lake:
\begin{eqnarray}
    \frac{\partial A}{\partial t} &= D \frac{\partial^2 A}{\partial x^2} + r \mu(N,x) \frac{A}{1 + V(A,x)} - c A^2;\\
    V(A,x) &= \frac{1}{1+A^{p(x)}}.
    \label{eq:lake_model}
\end{eqnarray}
Here, algae growth is modelled as logistic growth with competition coefficient $c$. The growth rate depends on the effect of nutrients $\mu(N,x)$ and the amount of vegetation $V(A)$, which is larger if there are few algae. The parameter $p$ represents the shallowness of the lake. More information, including parameter values and the rescaling process, can be found in~\ref{sec:app_lakes}.

Here, we have taken the shalllowness $p(x)$ and the nutrient effect on the growth rate $\mu(N,x)$ as space-dependent, and consider the (spatial) average nutrient effect as bifurcation parameter. An example bifurcation diagram is given in Figure~\ref{fig:BD-lakes}. This shows the possibility of coexistence states in such systems -- in particular when transitioning from a turbid state.

\section{Discussion} \label{sec:discussion}

Classically, the tipping behaviour in complex systems has been illustrated using simple conceptual models with two alternative states~\cite{lenton2013environmental, holling1973resilience, may1977thresholds, scheffer2001catastrophic}. In this study, we have investigated a prototypical tipping system and extended it by making it spatially explicit, via the incorporation of spatial transport and spatial heterogeneity. The resulting spatially extended model system revealed a more intricate bifurcation structure with multiple additional branches of stable states compared to the classic `S'-curve bifurcation diagram. This is due to the presence of stable coexistence states with part of the domain in one state and part in another. Further, it revealed the possibility of partial transition in which only part of the system undergoes reorganization, limiting the impact of these events on the full system's functioning. Therefore, this study further specifies how tipping of spatially heterogeneous systems happens; they do not necessarily tip fully in one event, but they might tip in smaller steps, each corresponding to a restructuring of part of the system only.

Partial tipping events prevent a full reorganisation of a system, instead giving rise to reorganisation of part of the system only. These events are therefore less severe than full tipping events, and lead to smaller hysteresis loops. On top of that, while part of the domain is still in its original state, restoration might also be easier and can happen more gradual: as climatic conditions would improve again, the spatial interface between states can move, slowly recovering the system.

This detail of the tipping process might be relevant for many natural systems. Here, we have illustrated this using a few conceptual example systems on different spatial scales, but there are many more. In principle, if a model or real system would allow for bistability locally, spatial effects can lead to the formation of coexistence states -- and thus these systems might experience partial tipping. One of the prominent examples is the occurrence of multiple convection patterns in ocean general circulation models \cite{Rahmstorf1995, Rahmstorf1994}. Such behavior was often attributed to numerical artifacts (due to the coarse horizontal resolution in such models). However, the different convection pattern can in the framework here be interpreted as different stable coexistence states. Indeed, the vertical coupling is very strong and follows basically the box model example in section \ref{sec:example_oceans}, whereas the horizontal diffusive coupling is rather weak. 

As illustrated, the precise importance and relevance of coexistence states and partial tipping depends a lot on the size of the system and the specific heterogeneity. For instance, if a system is too confined, coexistence states cannot form at all, and the spatial heterogeneities change the bifurcation structure and the severity of (partial) tipping events. Consequently, for some systems these partial tipping events might be more important than for others.

One of the most important tasks remaining is therefore to distinguish between the type and severity of a (partial) tipping event before it happens. Current generic early warning signs seem focused mostly on predicting when an imminent system change is going to happen~\cite{scheffer2009early, scheffer2012anticipating, lenton2011early}, but not on what is happening then. If it is only a minor partial reorganization, it does not need to be worrisome. Further, in spatially-extended systems early warning signs, like critical slowing down, might only be visible in the part of the domain in which change is going to happen, while it might be absent in the global response. Both of these issues might be tackled by inspection of the spatial structure of the destabilizing perturbation~\cite{bastiaansen2020effect}.

In this paper, we have deliberately used conceptual models that are very basic and relatively simple to showcase the general phenomenon and argue its generality. In more complex and realistic models, coexistence states and partial tipping events should also be present in general. However, dynamics of such models can be much more difficult. For instance, models with 2D or 3D spatial domains, the spatial interfaces can have complicated structures, and in systems with multiple components might also undergo bifurcations~\cite{bastiaansen2019stable, bel2012gradual, fernandez2019front}.

Finally, we have focused on stationary states, but also the transient behaviour of coexistence states is relevant. When not in equilibrium, a spatial interface slowly moves towards its equilibrium position, hereby converting part of the domain from one state to another -- but slowly~\cite{zelnik2018regime, bel2012gradual}. This contrast the (partial) tipping events, in which change is fast. If we would have a better understanding of this transient behaviour of the spatial interface, and how it is influenced by climate change and (human-made) heterogeneities, it might be possible to create circumstances that allow these systems to slowly restore themselves naturally more easily.

\newpage

\appendix

\section{Mathematical Theory of Coexistence States}\label{sec:app_math_coexistence}

In this appendix section, we briefly explain the mathematical ideas behind the presented coexistence states. This does not constitute a rigorous mathematical proof; many details, in particular in the spatially heterogeneous case, warrant a further, more careful mathematical investigation.

\subsection{Spatially homogeneous equation}

\begin{figure}[p]
    \centering

     \begin{subfigure}[t]{0.19\textwidth}
            \centering
            \includegraphics[width=\textwidth]{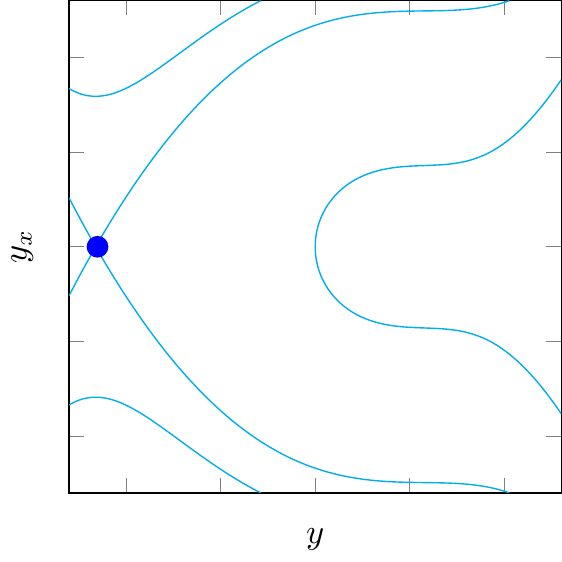}
            \caption{$\mu < \mu_B$}
    \end{subfigure}
    \begin{subfigure}[t]{0.19\textwidth}
            \centering
            \includegraphics[width=\textwidth]{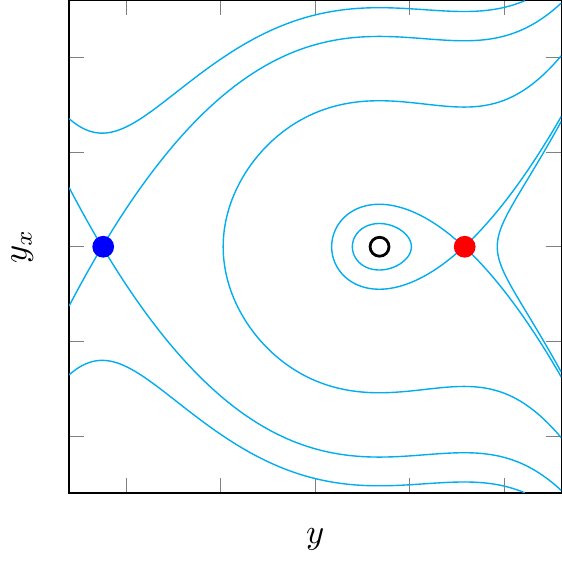}
            \caption{$\mu_B < \mu < \mu_M$}
    \end{subfigure}    
    \begin{subfigure}[t]{0.19\textwidth}
            \centering
            \includegraphics[width=\textwidth]{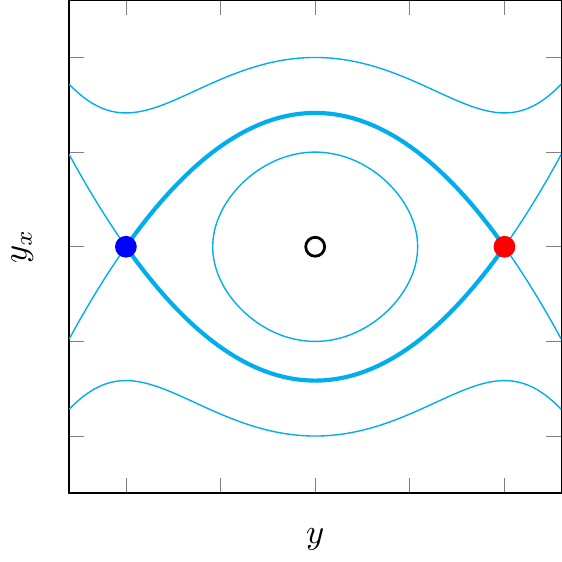}
            \caption{$\mu = \mu_M$}
    \end{subfigure}    
    \begin{subfigure}[t]{0.19\textwidth}
            \centering
            \includegraphics[width=\textwidth]{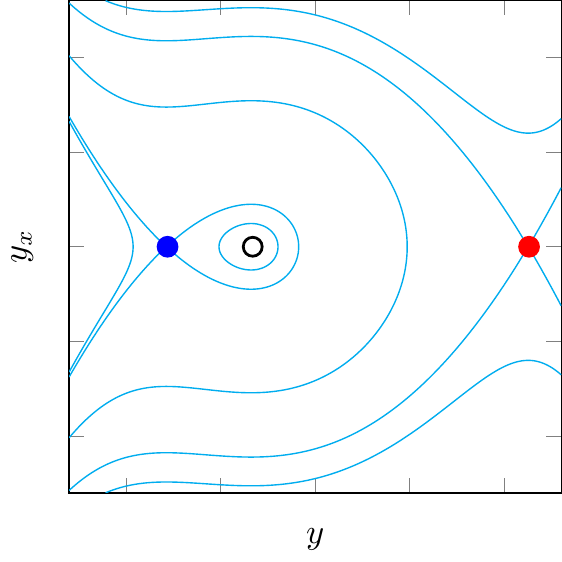}
            \caption{$\mu_M < \mu < \mu_A$}
    \end{subfigure}    
    \begin{subfigure}[t]{0.19\textwidth}
            \centering
            \includegraphics[width=\textwidth]{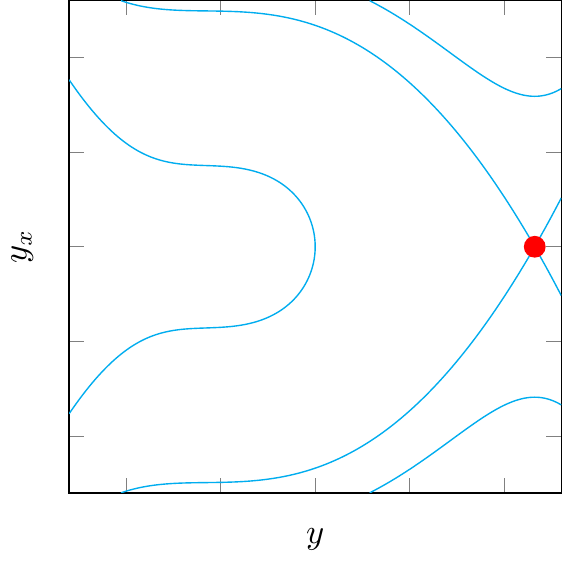}
            \caption{$\mu > \mu_A$}
    \end{subfigure}

    \caption{Level curves for a typical Hamiltonian system for various values of parameter $\mu$. The circles correspond to fixed points of~(\ref{eq:app_spatial_equilibrium}), and hence to spatially uniform solutions of ~(\ref{eq:spatial_model_diffusion}). Blue circles denote the lower stable state, red circles the higher stable state and black, non-filled circles the unstable uniform state. The thick cyan lines in (c) indicate the spatial interfaces (also called fronts, or heteroclinic connections) that are the building blocks of the coexistence states.}
    \label{fig:app_hamiltonian}
\end{figure}

Equilibrium solutions to (\ref{eq:spatial_model_diffusion}) (in one spatial dimension) have to satisfy
\begin{equation}
    0 = D y_{xx} + f(y; \mu),
    \label{eq:app_spatial_equilibrium}
\end{equation}
where the subscripts $x$ denote taking the derivative with respect to $x$. This is a Hamiltonian system with Hamiltonian $H$ given by
\begin{equation}
    H(y, y_x; \mu) = D \frac{y_x^2}{2} + V(y;\mu)
\end{equation}
where $V$ is the potential function, i.e. $V$ satisfies
\begin{equation}
    \frac{\partial V}{\partial y}(y;\mu) = f(y; \mu).
    \label{eq:app_definition_F}
\end{equation}
Solutions to (\ref{eq:app_spatial_equilibrium}) therefore lie on the level curves with constant $H(y,y_x; \mu)$. Depending on the value of the (bifurcation) parameter $\mu$, the level sets can trace out different orbits in the phase space; see Figure~\ref{fig:app_hamiltonian}.

Equilibrium solutions to (\ref{eq:app_spatial_equilibrium}) are represented in these phase portraits as bounded orbits. For $\mu < \mu_B$, the only bounded solution is the point $(y,y_x) = (y_A,0)$ corresponding to the uniform solution $y(x) \equiv y_A$ to (\ref{eq:app_spatial_equilibrium}). At $\mu = \mu_B$ a saddle-node bifurcation occurs which creates two other uniform solutions ($y(x) \equiv y_B$ and another, unstable one). For $\mu_B < \mu < \mu_M$, there are additional bounded solutions: a homoclinic orbit connecting $(y_B,0)$ to itself (corresponding to a pulse solution to (\ref{eq:app_spatial_equilibrium})) and periodic orbits, but these are all unstable solutions to the PDE. At $\mu = \mu_M$, the Maxwell point, the points $(y_A,0)$ and $(y_B,0)$ lie on the same level curve, i.e. $\Delta H(\mu) := H(y_A,0;\mu) - H(y_B,0;\mu) = V(y_A;\mu) - V(y_B;\mu) = 0$ for $\mu = \mu_M$, and heteroclinic connections between both points exist, which represent the stable equilibrium co-existence states to (\ref{eq:spatial_model_diffusion}). For $\mu_M < \mu < \mu_A$, there is a homoclinic orbit to $(y_A,0)$, and periodic orbits that are all unstable. Finally, for $\mu > \mu_A$ only the uniform solution $y(x) \equiv y_B$ exists.

\subsection{Spatially heterogeneous equation - Existence of equilibrium solutions}

\begin{figure}[p]
    \centering
    \includegraphics[width = 0.65\textwidth]{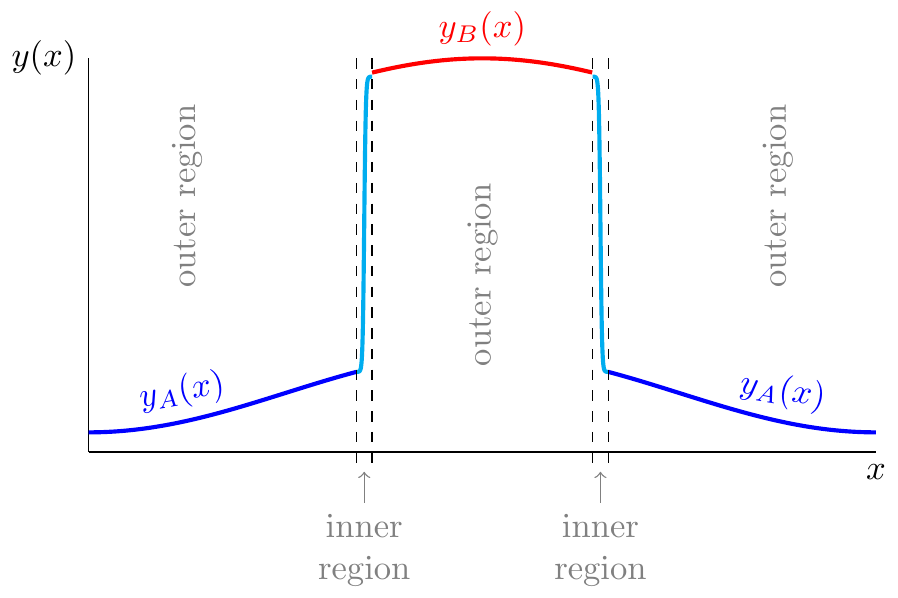}
    \caption{Sketch of a coexistence state in spatially heterogeneous models. Part of the domain resides in state A (blue), and part in state B (red), with a spatial interface (cyan) connecting these regions.}
    \label{fig:app_solution_form}
\end{figure}

Equilibrium solutions to (\ref{eq:spatial_model}) (in one spatial dimension) satisfy
\begin{equation}
    0 = D y_{xx} + f(y,x ;\mu)
    \label{eq:app_spatial_equilibrium_het}
\end{equation}
We set $D = \varepsilon^2 \ll 1$ (a change of coordinate $x$ can achieve this) and we assume that $\frac{\partial f}{\partial x}(y, x;\mu) \leq \mathcal{O}(1)$ (with respect to $\varepsilon$). In this setting, coexistence solutions to (\ref{eq:app_spatial_equilibrium_het}) have the form as depicted in Figure~\ref{fig:app_solution_form}: there are large, so-called outer regions that reside in the same state and small, so-called inner regions in which the transitions between regions are located (i.e. the spatial interfaces). Solutions to (\ref{eq:app_spatial_equilibrium_het}) can now be constructed per region, and should be matched at the boundaries. Here, we only give a sketch of the leading order construction per region. We note that a rigorous extension of the ideas formulated here would require higher-order terms and matching of the regions.

In the outer region, (\ref{eq:app_spatial_equilibrium_het}) becomes at leading order the equation
\begin{equation}
    0 = f(y, x; \mu),
\end{equation}
with the added constraint that $y$ needs to be smooth. Hence, this leads to $y(x) = y_A(x)$ or $y = y_B(x)$.

In the inner regions, we use the coordinate change $\xi := \frac{x - x_F}{\varepsilon}$, where we zoom in on the position $x = x_F$ where the interface is located. We then obtain
\begin{equation}
    0 = y_{\xi\xi} + f(y, x_F; \mu) + \varepsilon \frac{\partial f}{\partial x}(y, x_F; \mu) \xi + h.o.t.
    \label{eq:app_EX_inner_region}
\end{equation}
where $h.o.t.$ stands for the higher-order terms. At leading order this equation does not depend on space (for any given $x_F$). Hence, the theory of interfaces in the spatial homogeneous model applies. That is, by extending the definition of $V$ in (\ref{eq:app_definition_F}) to
\begin{equation}
    \frac{\partial V}{\partial y}(y, x; \mu) = f(y, x; \mu),
\end{equation}
and defining
\begin{equation}
    \Delta H(x, \mu) := V(y_A(x), x; \mu) - V(y_B(x), x; \mu),
\end{equation}
it is now clear that an interface can only be located at $x = x_F$ when $\Delta H(x_F,\mu) = 0$.

\subsection{Spatially heterogeneous equation - Stability of equilibrium solutions}

For any given stationary state $y = y^*$, the (linear) stability can be analyzed by setting $y(x,t) = y^*(x) + e^{\lambda t} \bar{y}(x)$ in  (\ref{eq:spatial_model}), which yields the linear stability problem
\begin{equation}
    \lambda \bar{y} = \varepsilon^2 \bar{y}_{xx} + \frac{\partial f}{\partial y}(y^*, x; \mu) \bar{y}
\end{equation}
Again, this equation can be analyzed per region.

In the outer regions we have approximately
\begin{equation}
    \lambda \bar{y} = \frac{\partial f}{\partial y}(y^*, x; \mu) \bar{y}.
\end{equation}
So an outer region contributes to the spectrum all values $\lambda = \frac{\partial f}{\partial y}(y^*(x), x; \mu)$ for all $x$ within the outer region. Heuristically, this means that the solution becomes unstable in an outer region if the local dynamics (so ignoring all spatial effects) would become unstable.

In the inner regions we have
\begin{equation}
    \lambda \bar{y} = \bar{y}_{\xi\xi} + \frac{\partial f}{\partial y}(y^*, x_F; \mu) \bar{y} + \varepsilon \frac{\partial^2 f}{\partial x \partial y}(y^*, x_F; \mu) \bar{y} \xi + h.o.t.
    \label{eq:app_STAB_inner_region}
\end{equation}
Upon differentiation of (\ref{eq:app_EX_inner_region}) with respect to $\xi$ we obtain
\begin{eqnarray}
    0  = &\ y^*_{\xi\xi\xi} + \frac{\partial f}{\partial y}(y^*,x_F;\mu) y^*_\xi + \varepsilon \frac{\partial^2 f}{\partial x \partial y}(y^*, x_F; \mu) y^*_\xi \xi \\ 
    &+ \varepsilon \frac{\partial f}{\partial x}(y^*; x_F; \mu) + h.o.t.
\end{eqnarray}
Therefore, substitution of $\bar{y} = y^*_\xi + \varepsilon \tilde{y}$ and $\lambda = \varepsilon \tilde{\lambda}$ into~(\ref{eq:app_STAB_inner_region}) yields at leading order the equation
\begin{equation}
    \tilde{y}_{\xi\xi} + \frac{\partial f}{\partial y}(y^*,x_f;\mu) \tilde{y} = \tilde{\lambda} y^*_\xi + \frac{\partial f}{\partial x}(y^*, x_F; \mu)
\end{equation}
A Fredholm solvability condition then indicates that each inner region contributes an eigenvalue
\begin{eqnarray}
    \lambda 
    & = - C \varepsilon \int_{I_f} \frac{\partial f}{\partial x}(y^*(\xi), x_F; \mu) y^*_\xi(\xi) d\xi \\
    & = - C \varepsilon \int_{I_f} \frac{d}{d\xi} \left( \frac{\partial V}{\partial x}(y^*(\xi), x_F; \mu) \right) d\xi \\
    & = - C \varepsilon \frac{\partial V}{\partial x}(y^*, x_F; \mu)|_{\partial I_f},
\end{eqnarray}
where $C > 0$ is a constant, and $I_f$ is the inner region.

Thus, $\lambda$ is proportional to the difference between $\frac{\partial V}{\partial x}$ at both ends of the interface. Or, put differently, to the derivative of $\Delta H$. If the interface under consideration goes from state $B$ to state $A$, we have $\lambda = - C\ \frac{\partial \Delta H}{\partial x}(x_F;\mu)$ and if it goes from $A$ to $B$ we have $\lambda = C\ \frac{\partial \Delta H}{\partial x}(x_F;\mu)$.

The equilibrium solution $y^*$ is only stable if all $\lambda$ in the spectrum have negative real parts. Therefore, this is only the case when (i) all outer regions are locally stable and (ii) all interfaces are located at positions that yield only negative eigenvalues $\lambda$.

\section{The small domain limit}\label{sec:small_domain_limit}
If the spatial domain is too small (compared to the diffusion strength), spatial interfaces do not fit in the domain and coexistence states cannot form. If the domain is really small, the spatial heterogeneity becomes irrelevant and dynamics are essentially similar to a model that has no spatial effects. To see this, consider the following spatially heterogeneous model with one spatial dimension:
\begin{equation}
    \frac{\partial y}{\partial t} = D \frac{\partial^2 y}{\partial x^2} + f(y, x; \mu)
\end{equation}
on the domain $[-L,L]$ with no-flux boundary conditions. By scaling $x \rightarrow x / L$, we obtain
\begin{equation}
    \frac{\partial y}{\partial t} = \frac{D}{L^2} \frac{\partial^2 y}{\partial x^2} + f(y, x;\mu)
\end{equation}
on the domain $[-1,1]$. Now, in the small domain limit, $\tilde{D} := \frac{D}{L^2} \gg 1$.

Hence, solutions can be approximated using perturbation techniques. Specifally, we can set
\begin{equation}
    y(x,t) = y_0(x,t) + \mathcal{O}\left(\frac{1}{\tilde{D}}\right).
\end{equation}
Then, at leading order,
\begin{equation}
    \frac{\partial^2 y_0}{\partial x^2} = 0
\end{equation}
which indicates that $y_0(x,t) \equiv y_0(t)$ is uniform in space.

Then, the evolution of the spatial average $\left<y\right>(t)$ can be found at leading order:
\begin{eqnarray}
    \frac{d}{dt} \left<y\right>(t) 
    & = \frac{1}{2} \int_{-1}^1 \frac{\partial y}{\partial t}(x,t)\ dx \\
    & = \frac{1}{2} \int_{-1}^1 \tilde{D}\frac{\partial^2 y}{\partial x^2}(x,t)\ dx + \frac{1}{2} \int_{-1}^1 f(y,x,\mu)\ dx \\
    & = \frac{1}{2} \int_{-1}^1 f(y_0,x;\mu)\ dx + \mathcal{O}\left(\frac{1}{\tilde{D}}\right),
\end{eqnarray}
where the integral over the spatial derivatives vanishes because of the no-flux boundary conditions. Since $y_0$ is uniform in space, the remaining integral is only the spatial average over the heterogeneity. That is,
\begin{equation}
    \frac{d}{dt} \left<y\right>(t) = F(\left<y\right>,t) := \frac{1}{2} \int_{-1}^{1} f(\left<y\right>,x;\mu)\ dx
\end{equation}
which has dynamics similar to a non-spatial model.

\section{Details of the 1D Global Energy Balance Model}\label{sec:app_GEBM}
The 1D global energy balance model in (\ref{eq:GEBM}) uses the following functional form for the temperature-dependent albedo:
\begin{equation}
    \alpha(T) = \alpha_1 + (\alpha_2-\alpha_1) \frac{1 + \tanh\left(K\left[T-\frac{T_1+T_2}{2}\right]\right)}{2},
\end{equation}
where $\alpha_1$ is the albedo of ice, and $\alpha_2$ the albedo of water. The sigmoid function ensures the change in albedo is smooth.

The incoming solar radiation is dependent on the latitude. Following~\cite{north1975analytical}, we have taken
\begin{equation}
    Q(x) = Q_0 \left(1 - 0.241\left[3x^2-1\right]\right),
\end{equation}
where $Q_0$ is the global average of the incoming solar radiation.

As parameters we have taken $Q_0 = 341.3 Wm^{-2}$, $\alpha_1 = 0.289$, $\alpha_2 = 0.70$, $T_1 = 260 K$, $T_2 = 290K$, $K = 0.1$, $\varepsilon = 0.61$, $\sigma_0 = 5.67 \cdot 10^{-8} W m^{-2} K^{-1}$. The value for the diffusivity $D$ is not known -- and potentially not uniform over earth~\cite{ghil1976climate, bodai2015global} -- but for the illustrative purposes in this paper we have taken $D = 0.30$. Finally, since we are focussing on stationary states, the value for $C_T$ does not matter, but a typical value would be $C_T = 5 \cdot 10^{8} J m^{-2} K^{-1}$.

\section{Details of the ocean convection model} \label{sec:app_ocean}
Equations (\ref{eq:ocean_model_eq1})-(\ref{eq:ocean_model_eq2}) describe the evolution of temperature $T$ and salinity $S$ in the ocean's mixed layer. We further simplify this two-component model as follows: density $\rho$ is approximated as a linear equation of state, i.e. $\rho = - \alpha T + \beta S$. Then, $T$ and $S$ are re-scaled as $\tilde{T} = \alpha T$, $\tilde{S} = \beta S$ such that $\rho = \tilde{S}-\tilde{T}$ and the system becomes
\begin{eqnarray}
    \frac{\partial \tilde{T}}{\partial t} & = D \frac{\partial^2 \tilde{T}}{\partial x^2} + k_T \left( \tilde{T}_A(x) - \tilde{T} \right) - \kappa(\Delta \rho) \left( \tilde{T} - \tilde{T}_0(x) \right);\\
        \frac{\partial \tilde{S}}{\partial t} & = D \frac{\partial^2 \tilde{S}}{\partial x^2} + k_S \left( \tilde{S}_A(x) - \tilde{S} \right) - \kappa(\Delta \rho) \left( \tilde{S} - \tilde{S}_0(x) \right).
\end{eqnarray}
Now, we assume $k_T = k_S$. Then, we define the spiciness $\mu := \tilde{S} + \tilde{T}$, and $\Delta \rho := \rho - \rho_0 = - (\tilde{T}-\tilde{T}_0)+(\tilde{S}-\tilde{S_0})$ and $\Delta \mu := \mu - \mu_0 = (\tilde{T}-\tilde{T}_0) + (\tilde{S}-\tilde{S}_0)$. The system can then be rewritten in terms of $\Delta \rho$ and $\Delta \mu$ as
\begin{eqnarray}
    \frac{\partial  \Delta \rho }{\partial t} & = D \frac{\partial^2 \Delta \rho }{\partial x^2} + k_T \left( \Delta \rho_A(x) - \Delta \rho \right) - \kappa(\Delta \rho) \Delta \rho + D \frac{\partial^2 \rho_0(x)}{\partial x^2} \\
    \frac{\partial \Delta \mu }{\partial t} & = D \frac{\partial^2  \Delta \mu }{\partial x^2} + k_T \left( \Delta \mu_A(x) - \Delta \mu \right) - \kappa(\Delta \rho) \Delta \mu + D \frac{\partial^2 \mu_0(x)}{\partial x^2}
\end{eqnarray}
Since the equation for $\Delta \rho$ does not depend on $\Delta \mu$, it can be solved without taking $\Delta \mu$ into account. Hence, for the purposes here, only the equation for $\Delta \rho$ is needed.

For the exchange rate $\kappa(\Delta \rho)$ we have taken the following functional form:
\begin{equation}
    \kappa(\Delta \rho) = \frac{\bar{\kappa}}{2} \left[ 1 + \tanh\left(\Delta \rho-\Delta \rho_\mathrm{ref}\right)\right]
\end{equation}
As parameter values we have taken $k_T = 1$, $\bar{\kappa} = 100$, $\Delta \rho_\mathrm{ref} = -1/2$, $\rho_0(x) \equiv 0$, $\rho_A(x; f) = 2 + f \left(1 + \cos(x \pi / 2) \right)$

\section{Details of the tropical forest model} \label{sec:app_tropicalforest}

The tropical forest model in (\ref{eq:tropical_forest}) uses the following functional forms, that have been taken from~\cite{wuyts2019tropical}:
\begin{eqnarray}
    r(P) & = 0.20 \left(1 - e^{1.54 - 0.003 P}\right); \\
    m(P) & = 0.041 + e^{-2.15 - 0.008 P}; \\
    f(F;P) & = \frac{0.46}{2.7} \frac{Y_c(P)^4}{Y_c(P)^4+F^4},
\end{eqnarray}
where $Y_c(P)$ models the assumed decrease in fire-spreading percolation threshold (see~\cite{staver2012integrating} for details), expressed in equations as
\begin{equation}
    Y_c(P) = \max\left( 0, 0.56 - 1.43 \cdot 10^{-4} P \right).
\end{equation}

In~\cite{wuyts2019tropical}, a 1D spatial domain $[0,3000 km]$ is taken with diffusivity $D = 0.2$. By rescaling space (see section~\ref{sec:small_domain_limit}) this is equivalent to a domain $[-1,1]$ with $D \approx 5 \cdot 10^{-6}$. However, we take $D = 10^{-4}$, essentially modelling a smaller real domain of about $100 km$.

Following~\cite{wuyts2019tropical} we assume a linear precipitation gradient over space. On top of that, we model climate change via a spatially uniform term. Specifically,
\begin{equation}
    P(x) = P_\mathrm{mean} + 150 x.
\end{equation}
The parameter $P_\mathrm{mean}$ is used as a bifurcation parameter in this example system.

\section{Details of the shallow lake model}\label{sec:app_lakes}

The shallow lake model that has been created in~\cite{scheffer1993alternative} is non-spatial and dimensional. It is given by the following ordinary differential equation
\begin{equation}
    \frac{dA}{dt} = r \left(\frac{N}{N+h_N}\right)\left(\frac{h_v}{h_v+V(A)}\right)A - c A^2,
\end{equation}
where
\begin{equation}
    V(A) = \frac{h_A^p}{h_A^p+A^p}.
\end{equation}
We define $\hat{\mu} := \frac{N}{N+h_N}$ and rescale $\hat{A} = A / h_A$, $\hat{c} = h_A c$ to obtain the scaled model
\begin{equation}
    \frac{dA}{dt} = r \hat{\mu} \frac{1}{1+\hat{V}(\hat{A})}\hat{A} - \hat{c}\hat{A}^2,
\end{equation}
where
\begin{equation}
    \hat{V}(\hat{A}) = \frac{1}{h_V} \frac{1}{1+\bar{A}^p}.
\end{equation}
With spatial effects added (and dropping the hats), this becomes the model in (\ref{eq:lake_model}).

For the numerical computations, we have taken the following space-dependent functions
\begin{eqnarray}
    \hat{\mu}(x) & = \mu \left( 1 - \frac{1}{10} \cos(\pi x) \right);\\
    p(x) & = 3 + \frac{1}{10} \cos(\pi x).
\end{eqnarray}
The other parameter values are $r = 10$, $h_v = 0.1$ and $c = 1$. The rescaled diffusion coefficient was taken as $D = 0.01$ (with spatial domain $[-1,1]$).

\newpage

\section*{Acknowledgements}
We thank Valerie Engelmayer for bringing to our attention the coexistence states in stratocumulus clouds. This project is TiPES contribution \#137: This project has received funding from the European Union’s Horizon 2020 research and innovation programme under grant agreement 820970.

\vspace{3em}

\bibliographystyle{unsrt}
\bibliography{sources}

\begin{thebibliography}{10}

\bibitem{lenton2013environmental}
Timothy~M Lenton.
\newblock Environmental tipping points.
\newblock {\em Annual Review of Environment and Resources}, 38:1--29, 2013.

\bibitem{scheffer2001catastrophic}
Marten Scheffer, Steve Carpenter, Jonathan~A Foley, Carl Folke, and Brian
  Walker.
\newblock Catastrophic shifts in ecosystems.
\newblock {\em Nature}, 413(6856):591--596, 2001.

\bibitem{holling1973resilience}
Crawford~S Holling.
\newblock Resilience and stability of ecological systems.
\newblock {\em Annual review of ecology and systematics}, 4(1):1--23, 1973.

\bibitem{drijfhout2015catalogue}
Sybren Drijfhout, Sebastian Bathiany, Claudie Beaulieu, Victor Brovkin, Martin
  Claussen, Chris Huntingford, Marten Scheffer, Giovanni Sgubin, and Didier
  Swingedouw.
\newblock Catalogue of abrupt shifts in {I}ntergovernmental {P}anel on
  {C}limate {C}hange climate models.
\newblock {\em Proceedings of the National Academy of Sciences},
  112(43):E5777--E5786, 2015.

\bibitem{lenton2008tipping}
Timothy~M Lenton, Hermann Held, Elmar Kriegler, Jim~W Hall, Wolfgang Lucht,
  Stefan Rahmstorf, and Hans~Joachim Schellnhuber.
\newblock Tipping elements in the {E}arth's climate system.
\newblock {\em Proceedings of the national Academy of Sciences},
  105(6):1786--1793, 2008.

\bibitem{hirota2011global}
Marina Hirota, Milena Holmgren, Egbert~H Van~Nes, and Marten Scheffer.
\newblock Global resilience of tropical forest and savanna to critical
  transitions.
\newblock {\em Science}, 334(6053):232--235, 2011.

\bibitem{staver2011global}
A~Carla Staver, Sally Archibald, and Simon~A Levin.
\newblock The global extent and determinants of savanna and forest as
  alternative biome states.
\newblock {\em science}, 334(6053):230--232, 2011.

\bibitem{rietkerk1997alternate}
Max Rietkerk and Johan van~de Koppel.
\newblock Alternate stable states and threshold effects in semi-arid grazing
  systems.
\newblock {\em Oikos}, pages 69--76, 1997.

\bibitem{rietkerk1997site}
Max Rietkerk, Frank van~den Bosch, and Johan van~de Koppel.
\newblock Site-specific properties and irreversible vegetation changes in
  semi-arid grazing systems.
\newblock {\em Oikos}, pages 241--252, 1997.

\bibitem{stocker1991rapid}
Thomas~F Stocker and Daniel~G Wright.
\newblock Rapid transitions of the ocean's deep circulation induced by changes
  in surface water fluxes.
\newblock {\em Nature}, 351(6329):729--732, 1991.

\bibitem{lohmann2021risk}
Johannes Lohmann and Peter~D Ditlevsen.
\newblock Risk of tipping the overturning circulation due to increasing rates
  of ice melt.
\newblock {\em Proceedings of the National Academy of Sciences}, 118(9), 2021.

\bibitem{garbe2020hysteresis}
Julius Garbe, Torsten Albrecht, Anders Levermann, Jonathan~F Donges, and
  Ricarda Winkelmann.
\newblock The hysteresis of the {A}ntarctic ice sheet.
\newblock {\em Nature}, 585(7826):538--544, 2020.

\bibitem{pattyn2020uncertain}
Frank Pattyn and Mathieu Morlighem.
\newblock The uncertain future of the antarctic ice sheet.
\newblock {\em Science}, 367(6484):1331--1335, 2020.

\bibitem{huybrechts1999dynamic}
Philippe Huybrechts and Jan De~Wolde.
\newblock The dynamic response of the greenland and {A}ntarctic ice sheets to
  multiple-century climatic warming.
\newblock {\em Journal of Climate}, 12(8):2169--2188, 1999.

\bibitem{scheffer1993alternative}
Marten Scheffer, S~Harry Hosper, Marie~Louise Meijer, Brian Moss, and Erik
  Jeppesen.
\newblock Alternative equilibria in shallow lakes.
\newblock {\em Trends in ecology \& evolution}, 8(8):275--279, 1993.

\bibitem{lenton2013origin}
Timothy~M Lenton and Hywel~TP Williams.
\newblock On the origin of planetary-scale tipping points.
\newblock {\em Trends in Ecology \& Evolution}, 28(7):380--382, 2013.

\bibitem{rockstrom2009safe}
Johan Rockstr{\"o}m, Will Steffen, Kevin Noone, {\AA}sa Persson, F~Stuart
  Chapin, Eric~F Lambin, Timothy~M Lenton, Marten Scheffer, Carl Folke,
  Hans~Joachim Schellnhuber, et~al.
\newblock A safe operating space for humanity.
\newblock {\em nature}, 461(7263):472--475, 2009.

\bibitem{steffen2018trajectories}
Will Steffen, Johan Rockstr{\"o}m, Katherine Richardson, Timothy~M Lenton, Carl
  Folke, Diana Liverman, Colin~P Summerhayes, Anthony~D Barnosky, Sarah~E
  Cornell, Michel Crucifix, et~al.
\newblock Trajectories of the {E}arth system in the {A}nthropocene.
\newblock {\em Proceedings of the National Academy of Sciences},
  115(33):8252--8259, 2018.

\bibitem{barnosky2012approaching}
Anthony~D Barnosky, Elizabeth~A Hadly, Jordi Bascompte, Eric~L Berlow, James~H
  Brown, Mikael Fortelius, Wayne~M Getz, John Harte, Alan Hastings, Pablo~A
  Marquet, et~al.
\newblock Approaching a state shift in {E}arth’s biosphere.
\newblock {\em Nature}, 486(7401):52--58, 2012.

\bibitem{may1977thresholds}
Robert~M May.
\newblock Thresholds and breakpoints in ecosystems with a multiplicity of
  stable states.
\newblock {\em Nature}, 269(5628):471--477, 1977.

\bibitem{vannes2005}
Egbert~H van Nes and Marten Scheffer.
\newblock Implications of spatial heterogeneity for catastrophic regime shifts
  in ecosystems.
\newblock {\em Ecology}, 86(7):1797--1807, 2005.

\bibitem{rosier2021tipping}
Sebastian~HR Rosier, Ronja Reese, Jonathan~F Donges, Jan De~Rydt, G~Hilmar
  Gudmundsson, and Ricarda Winkelmann.
\newblock The tipping points and early warning indicators for {Pine Island
  Glacier, West Antarctica}.
\newblock {\em The Cryosphere}, 15(3):1501--1516, 2021.

\bibitem{valdes2011built}
Paul Valdes.
\newblock Built for stability.
\newblock {\em Nature Geoscience}, 4(7):414--416, 2011.

\bibitem{rietkerk2021evasion}
Max Rietkerk, Robbin Bastiaansen, Swarnendu Banerjee, Johan van~de Koppel, Mara
  Baudena, and Arjen Doelman.
\newblock Evasion of tipping in complex systems through spatial pattern
  formation.
\newblock {\em Science}, 374(6564):eabj0359, 2021.

\bibitem{bastiaansen2020effect}
Robbin Bastiaansen, Arjen Doelman, Maarten~B Eppinga, and Max Rietkerk.
\newblock The effect of climate change on the resilience of ecosystems with
  adaptive spatial pattern formation.
\newblock {\em Ecology letters}, 23(3):414--429, 2020.

\bibitem{gildor2001physical}
Hezi Gildor and Eli Tziperman.
\newblock Physical mechanisms behind biogeochemical glacial-interglacial co2
  variations.
\newblock {\em Geophysical Research Letters}, 28(12):2421--2424, 2001.

\bibitem{alkhayuon2019basin}
Hassan Alkhayuon, Peter Ashwin, Laura~C Jackson, Courtney Quinn, and Richard~A
  Wood.
\newblock Basin bifurcations, oscillatory instability and rate-induced
  thresholds for {A}tlantic meridional overturning circulation in a global
  oceanic box model.
\newblock {\em Proceedings of the Royal Society A}, 475(2225):20190051, 2019.

\bibitem{scheffer2012anticipating}
Marten Scheffer, Stephen~R Carpenter, Timothy~M Lenton, Jordi Bascompte,
  William Brock, Vasilis Dakos, Johan Van~de Koppel, Ingrid~A Van~de Leemput,
  Simon~A Levin, Egbert~H Van~Nes, et~al.
\newblock Anticipating critical transitions.
\newblock {\em science}, 338(6105):344--348, 2012.

\bibitem{scheffer2003catastrophic}
Marten Scheffer and Stephen~R Carpenter.
\newblock Catastrophic regime shifts in ecosystems: linking theory to
  observation.
\newblock {\em Trends in ecology \& evolution}, 18(12):648--656, 2003.

\bibitem{Micke2018}
Kevin Micke.
\newblock Every pixel of {GOES-17} imagery at your fingertips.
\newblock {\em Bulletin of the American Meteorological Society}, 99(11):2217 --
  2219, 2018.

\bibitem{allen1972ground}
Samuel~Miller Allen and John~W Cahn.
\newblock Ground state structures in ordered binary alloys with second neighbor
  interactions.
\newblock {\em Acta Metallurgica}, 20(3):423--433, 1972.

\bibitem{bray2002theory}
Alan~J Bray.
\newblock Theory of phase-ordering kinetics.
\newblock {\em Advances in Physics}, 51(2):481--587, 2002.

\bibitem{carr1989metastable}
Jack Carr and Robert~L Pego.
\newblock Metastable patterns in solutions of $u_t= \varepsilon^2u_{xx}- f
  (u)$.
\newblock {\em Communications on pure and applied mathematics}, 42(5):523--576,
  1989.

\bibitem{sandstede2002stability}
Bj{\"o}rn Sandstede.
\newblock Stability of travelling waves.
\newblock In {\em Handbook of dynamical systems}, volume~2, pages 983--1055.
  Elsevier, 2002.

\bibitem{pismen2006patterns}
Len~M Pismen.
\newblock {\em Patterns and interfaces in dissipative dynamics}.
\newblock Springer Science \& Business Media, 2006.

\bibitem{zelnik2018regime}
Yuval~R Zelnik and Ehud Meron.
\newblock Regime shifts by front dynamics.
\newblock {\em Ecological Indicators}, 94:544--552, 2018.

\bibitem{bel2012gradual}
Golan Bel, Aric Hagberg, and Ehud Meron.
\newblock Gradual regime shifts in spatially extended ecosystems.
\newblock {\em Theoretical Ecology}, 5(4):591--604, 2012.

\bibitem{budyko1969effect}
Mikhail~I Budyko.
\newblock The effect of solar radiation variations on the climate of the
  {E}arth.
\newblock {\em Tellus}, 21(5):611--619, 1969.

\bibitem{sellers1969global}
William~D Sellers.
\newblock A global climatic model based on the energy balance of the
  {E}arth-atmosphere system.
\newblock {\em Journal of Applied Meteorology}, 8(3):392--400, 1969.

\bibitem{ikeda1999study}
Takashi Ikeda and Eiichi Tajika.
\newblock A study of the energy balance climate model with {CO2}-dependent
  outgoing radiation: Implication for the glaciation during the {C}enozoic.
\newblock {\em Geophysical Research Letters}, 26(3):349--352, 1999.

\bibitem{north1981energy}
Gerald~R North, Robert~F Cahalan, and James~A Coakley~Jr.
\newblock Energy balance climate models.
\newblock {\em Reviews of Geophysics}, 19(1):91--121, 1981.

\bibitem{ghil1976climate}
Michael Ghil.
\newblock Climate stability for a {S}ellers-type model.
\newblock {\em Journal of Atmospheric Sciences}, 33(1):3--20, 1976.

\bibitem{bodai2015global}
Tam{\'a}s B{\'o}dai, Valerio Lucarini, Frank Lunkeit, and Robert Boschi.
\newblock Global instability in the {G}hil--{S}ellers model.
\newblock {\em Climate Dynamics}, 44(11-12):3361--3381, 2015.

\bibitem{thorndike2012multiple}
Alan Thorndike.
\newblock Multiple equilibria in a minimal climate model.
\newblock {\em Cold regions science and technology}, 76:3--7, 2012.

\bibitem{north1984small}
Gerald~R North.
\newblock The small ice cap instability in diffusive climate models.
\newblock {\em Journal of Atmospheric Sciences}, 41(23):3390--3395, 1984.

\bibitem{north1975analytical}
Gerald~R North.
\newblock Analytical solution to a simple climate model with diffusive heat
  transport.
\newblock {\em Journal of Atmospheric Sciences}, 32(7):1301--1307, 1975.

\bibitem{drazin1977branching}
PG~Drazin and DH~Griffel.
\newblock On the branching structure of diffusive climatological models.
\newblock {\em Journal of Atmospheric Sciences}, 34(11):1696--1706, 1977.

\bibitem{caldeira1992susceptibility}
Ken Caldeira and James~F Kasting.
\newblock Susceptibility of the early {E}arth to irreversible glaciation caused
  by carbon dioxide clouds.
\newblock {\em Nature}, 359(6392):226--228, 1992.

\bibitem{hoffman2002snowball}
Paul~F Hoffman and Daniel~P Schrag.
\newblock The snowball {E}arth hypothesis: testing the limits of global change.
\newblock {\em Terra nova}, 14(3):129--155, 2002.

\bibitem{ashwin2020extreme}
Peter Ashwin and S~Anna.
\newblock Extreme sensitivity and climate tipping points.
\newblock {\em Journal of Statistical Physics}, 179(5):1531--1552, 2020.

\bibitem{welander1982simple}
Pierre Welander.
\newblock A simple heat-salt oscillator.
\newblock {\em Dynamics of Atmospheres and Oceans}, 6(4):233--242, 1982.

\bibitem{lenderink1994variability}
G~Lenderink and RJ~Haarsma.
\newblock Variability and multiple equilibria of the thermohaline circulation
  associated with deep-water formation.
\newblock {\em Journal of Physical Oceanography}, 24(7):1480--1493, 1994.

\bibitem{den2011spurious}
Matthijs den Toom, Henk~A Dijkstra, and Fred~W Wubs.
\newblock Spurious multiple equilibria introduced by convective adjustment.
\newblock {\em Ocean Modelling}, 38(1-2):126--137, 2011.

\bibitem{vellinga1998multiple}
Michael Vellinga.
\newblock Multiple equilibria in ocean models as a side effect of convective
  adjustment.
\newblock {\em Journal of physical oceanography}, 28(4):621--633, 1998.

\bibitem{kaiser2020detecting}
Amandine Kaiser, Davide Faranda, Sebastian Krumscheid, Danijel
  Belu{\v{s}}i{\'c}, and Nikki Vercauteren.
\newblock Detecting regime transitions of the nocturnal and {P}olar
  near-surface temperature inversion.
\newblock {\em Journal of the Atmospheric Sciences}, 77(8):2921--2940, 2020.

\bibitem{vandewiel2017regime}
Bas~JH Van~de Wiel, Etienne Vignon, Peter Baas, Ivo~GS Van~Hooijdonk, Steven~JA
  Van~der Linden, J~Antoon~van Hooft, Fred~C Bosveld, Stefan~R de~Roode,
  Arnold~F Moene, and Christophe Genthon.
\newblock Regime transitions in near-surface temperature inversions: A
  conceptual model.
\newblock {\em Journal of the Atmospheric Sciences}, 74(4):1057--1073, 2017.

\bibitem{rietkerk2004self}
Max Rietkerk, Stefan~C Dekker, Peter~C De~Ruiter, and Johan van~de Koppel.
\newblock Self-organized patchiness and catastrophic shifts in ecosystems.
\newblock {\em Science}, 305(5692):1926--1929, 2004.

\bibitem{aleman2020floristic}
JC~Aleman, Adeline Fayolle, Charly Favier, Ann~Carla Staver, Kyle~Graham
  Dexter, CM~Ryan, Akomian~Fortun{\'e} Azihou, David Bauman, Mariska te~Beest,
  Emmanuel~Ngulube Chidumayo, et~al.
\newblock Floristic evidence for alternative biome states in tropical {A}frica.
\newblock {\em Proceedings of the National Academy of Sciences},
  117(45):28183--28190, 2020.

\bibitem{bastiaansen2019stable}
Robbin Bastiaansen, Paul Carter, and Arjen Doelman.
\newblock Stable planar vegetation stripe patterns on sloped terrain in dryland
  ecosystems.
\newblock {\em Nonlinearity}, 32(8):2759, 2019.

\bibitem{bastiaansen2020pulse}
Robbin Bastiaansen, Martina Chirilus-Bruckner, and Arjen Doelman.
\newblock Pulse solutions for an extended {K}lausmeier model with spatially
  varying coefficients.
\newblock {\em SIAM Journal on Applied Dynamical Systems}, 19(1):1--57, 2020.

\bibitem{gandhi2018topographic}
Punit Gandhi, Lucien Werner, Sarah Iams, Karna Gowda, and Mary Silber.
\newblock A topographic mechanism for arcing of dryland vegetation bands.
\newblock {\em Journal of The Royal Society Interface}, 15(147):20180508, 2018.

\bibitem{wuyts2019tropical}
Bert Wuyts, Alan~R Champneys, Nicolas Verschueren, and Jo~I House.
\newblock Tropical tree cover in a heterogeneous environment: A
  reaction-diffusion model.
\newblock {\em PloS one}, 14(6):e0218151, 2019.

\bibitem{staver2012integrating}
A~Carla Staver and Simon~A Levin.
\newblock Integrating theoretical climate and fire effects on savanna and
  forest systems.
\newblock {\em The American Naturalist}, 180(2):211--224, 2012.

\bibitem{Rahmstorf1995}
S.~Rahmstorf.
\newblock {Multiple convection patterns and thermohaline flow in an idealized
  OGCM}.
\newblock {\em J Climate}, 8:3028--3039, 1995.

\bibitem{Rahmstorf1994}
S.~Rahmstorf.
\newblock {Rapid climate transitions in a coupled ocean-atmosphere model}.
\newblock {\em Nature}, 372:82--84, 1994.

\bibitem{scheffer2009early}
Marten Scheffer, Jordi Bascompte, William~A Brock, Victor Brovkin, Stephen~R
  Carpenter, Vasilis Dakos, Hermann Held, Egbert~H Van~Nes, Max Rietkerk, and
  George Sugihara.
\newblock Early-warning signals for critical transitions.
\newblock {\em Nature}, 461(7260):53--59, 2009.

\bibitem{lenton2011early}
Timothy~M Lenton.
\newblock Early warning of climate tipping points.
\newblock {\em Nature climate change}, 1(4):201--209, 2011.

\bibitem{fernandez2019front}
Cristian Fernandez-Oto, Omer Tzuk, and Ehud Meron.
\newblock Front instabilities can reverse desertification.
\newblock {\em Physical review letters}, 122(4):048101, 2019.

\end{thebibliography}

\end{document}